# Imaging Coulomb interactions and migrating Dirac cones in twisted graphene by local quantum oscillations

Matan Bocarsly[1†], Indranil Roy[1†], Vishal Bhardwaj[1†], Matan Uzan[1], Patrick Ledwith[2], Gal Shavit[3,4], Nasrin Banu[1], Yaozhang Zhou[1], Yuri Myasoedov[1], Kenji Watanabe[5], Takashi Taniguchi[6], Yuval Oreg[1], Dan Parker[7], Yuval Ronen[1], and Eli Zeldov[1]*

**Abstract**:

Flat band moiré graphene systems have emerged as a quintessential platform to investigate correlated phases of matter. A plethora of interaction-driven ground states have been proposed, and yet despite extensive experimental effort, there has been little direct evidence that distinguishes between the various phases, in particular near charge neutrality point. Here, we use a nanoscale scanning superconducting quantum interference device to image the local thermodynamic quantum oscillations in alternating-twist trilayer graphene at magnetic fields as low as 56 mT, which reveal ultrafine details of the density of states and of the renormalization of the single-particle band structure by Coulomb interactions. We find that the charging self-energy due to occupied electronic states, is critical in explaining the high carrier density physics. At half-filling of the conduction flat band, we observe a Stoner-like symmetry breaking, suggesting that it is the most robust mechanism in the hierarchy of phase transitions. On approaching charge neutrality, where the charging energy is negligible and exchange energy is dominant, we find the ground state to be a nematic semimetal which is favored over gapped states in the presence of heterostrain. In the revealed semimetallic phase, the flat-band Dirac cones migrate towards the mini-Brillouin zone center, spontaneously breaking the $C_3$ rotational symmetry. Our low-field local quantum oscillations technique presents an alluring avenue to explore the ground states of diverse strongly interacting van der Waals systems.

[1]Department of Condensed Matter Physics, Weizmann Institute of Science, Rehovot 7610001, Israel

[2]Department of Physics, Harvard University, Cambridge, MA 02138, USA

[3]Department of Physics and Institute for Quantum Information and Matter, California Institute of Technology, Pasadena, California 91125, USA

[4]Walter Burke Institute of Theoretical Physics, California Institute of Technology, Pasadena, California 91125, USA

[5]Research Center for Electronic and Optical Materials, National Institute for Materials Science; 1-1 Namiki, Tsukuba 305-0044, Japan

[6]Research Center for Materials Nanoarchitectonics, National Institute for Materials Science; 1-1 Namiki, Tsukuba 305-0044, Japan

[7]Department of Physics, University of California at San Diego, La Jolla, California 92093, USA

[†]These authors contributed equally to this work

*eli.zeldov@weizmann.ac.il

Moiré materials offer a fantastic combination of strongly correlated phases together with high experimental tunability, providing a controllable window into the mechanisms and origins of interacting quantum materials. The example of magic angle twisted bilayer graphene (MATBG) [1–10] led the way, with flat bands whose electronic interactions produce prominent correlated insulating states and unconventional superconductivity. Subsequently, the same basic phenomena were discovered in alternating-twist trilayer graphene (tTLG) [11–18], where the top and bottom layers are twisted with the same angle $\theta$ relative to the middle layer (Fig. 1a). This similarity is a consequence of mirror symmetry, which splits the tTLG band structure into two decoupled sectors [19,20]: (i) MATBG-like flat bands (FBs) that occur at a magic angle that is $\sqrt{2}$ larger than that of the MATBG, and (ii) a "bystander" monolayer-graphene-like Dirac cone. The presence of the extra decoupled Dirac cone sector in the tTLG band structure has two important gains over MATBG. First, the Dirac cone sector may be utilized as an effective probe of the correlated physics in the flat-band sector [17]. Second, application of a transverse displacement field $D$ breaks the mirror symmetry and hybridizes the two sectors, thus offering enlarged phase space for rich many-body physics and providing a new degree of in situ tunability. Indeed, possibly due to the larger magic angle, the correlated insulators and superconductors in tTLG are often more stable than in MATBG [21], making it a prime platform to explore the nature of these phases.

However, despite extensive theoretical proposals and vigorous experimental efforts, the identity of many phases — including the central phase at charge neutrality — remains unknown. Theoretical studies [22–36], including self-consistent Hartree-Fock (HF) and strong coupling analytics, considered symmetry broken states of magic angle FB systems at integer fillings, where the filling $-4 \leq \nu \leq 4$ of the eight flat bands is measured relative to charge neutrality. A wide variety of ground states have been proposed, including a valley polarized (VP) state, valley Hall (VH) order, and a Kramer's intervalley coherent (KIVC) state [23], all of which are gapped, and a gapless nematic semimetal (NSM) that spontaneously breaks the $C_3$ symmetry [30]. However, there is yet little direct experimental evidence distinguishing between these orders, especially at charge neutrality. Transport studies have revealed Chern insulators at some of the integer fillings [3,8,10], but still many of the underlying symmetries of the various many-body states remain unclear. More broadly, transport measurements are not thermodynamic and do not directly probe the density of states (DOS), making them less suited to identify ground states. In contrast, merely a few techniques directly probe the band structure (BS). One example is the scanning tunneling microscopy (STM) measurements that have quite recently revealed an incommensurate Kekulé spiral (IKS) order at $\nu = \pm 2$ in twisted graphene systems [18,37,38]. From a theoretical perspective, the ground state at the charge neutrality point (CNP) plays a central role in determining the entire phase diagram. Yet, barring observations of $C_3$ symmetry breaking by STM studies in some systems [18,39–42], there remains little experimental exploration of CNP ground states. This provides strong motivation to develop experimental techniques with the capability to probe charge neutrality.

Here we study the local thermodynamic de Haas-van Alphen (dHvA) oscillations [43,44] in tTLG slightly away from the magic angle. We detect quantum oscillations (QOs) in the Dirac sector down to remarkably low magnetic fields of 56 mT. This is facilitated by the low DOS of the Dirac sector and the lower sensitivity of the local measurements to the effects of disorder. In addition to providing high-energy resolution, these oscillations allow us to probe the ground state physics and to directly visualize the symmetry breaking and band structure renormalization due to the direct and exchange Coulomb interactions in the FBs.

By employing a self-consistent mean field theoretical study, we find that the flat-band dispersion is substantially renormalized due to the Coulomb repulsion. The Hartree interaction term, which describes the charging self-energy as a result of the background electronic charge in the system, is solely needed to describe the high-density physics, such as the evolution of the bandwidth with filling [15,45–47]. At half-filling of the conduction band we find clear evidence of spontaneous flavor-symmetry breaking in the flat-band sector [6,48]. The presence of this symmetry-broken generalized Stoner phase in the non-magic-angle regime, along with the absence of additional flavor polarization transitions or correlated insulators, strongly hints at a hierarchy of symmetry breaking in moiré graphene. Our findings suggest that the Stoner polarization near half-filling is most robust of these transitions, directly establishing the symmetry broken phase as a parent state of the pervasive $\nu = 2$ correlated insulator in MATBG [49,50].

Most importantly, near charge neutrality we find the exchange interaction, modeled by the mean field Fock term, is vital in understanding the effects of the displacement field $D$. We observe direct signatures of a FB whose Dirac cones have migrated from the Brillouin zone corners, and find strong evidence that the CNP ground state is the NSM which breaks $C_3$ symmetry. This phase is not generally theoretically favored in the presence of strong interactions [23], however even a small amount of heterostrain can make it energetically favorable [51]. Our findings of a semimetallic ground state provide invaluable insight into the long-standing puzzle of absence of observable gap at CNP in transport measurements in MATBG.

**Transport measurements**

The tTLG sample was fabricated using the dry-transfer method and encapsulated in hBN (Figs. 1a,b, Methods). The $dc$ voltages $V_{tg}^{dc}$ and $V_{bg}^{dc}$ applied to the top and bottom gates allow for control of the carrier density $n$ and out-of-plane displacement field $D$. Transport measurements of $R_{xx}$ and $R_{yx}$ at constant $D = 0$, were performed at a temperature $T = 300$ mK as a function of applied out-of-plane magnetic field $B_a$ and $n$ (Figs. 1c,d). At $D = 0$, the mirror symmetry allows the Hamiltonian of tTLG to be decomposed into a FB (MATBG) sector and a decoupled Dirac cone (Fig. 2h) [19,20]. As such, two Landau fans emanate from CNP; a dense fan emerging at large $B_a$ reflecting Landau levels (LLs) in the FBs, and a highly dispersing fan at low $B_a$ reflecting the LLs arising from the Dirac sector (Fig. 1c and Extended Data Fig. 1a). As the Dirac sector is populated in parallel with the FBs, it acts to shunt any gapped state in the FB, including the fourfold filling of the FBs (which is highly resistive in MATBG). Therefore, extracting information from transport demands a careful analysis of the two sectors [17].

The Chern number sequence in the Dirac band is $C_D = 4N_D + 2$, and the energies of the corresponding LLs are,

$$\varepsilon_N^D = sgn(N_D)v_F\sqrt{2e\hbar|N_D||B_a}.$$ (1)

Here, $e$ is the elementary charge, $\hbar$ the reduced Planck constant, $v_F$ is the Fermi velocity of the Dirac cone, and $N_D = 0, \pm1, \pm2,...$ is the Landau level index in the Dirac sector. The $N_D = 0$ LL resides at CNP with $C_D = 2$, and the half-filled compressible $N_D = 1$ LL (dotted line in Fig. 1c) crosses the edge of the FB at $n_0 = 4.03 \times 10^{12}$ cm$^{-2}$ and $B_a = 2.25$ T (blue circle). At this point in phase space, the $N_D = 0$ LL is filled along with half of the $N_D = 1$ LL. Therefore, the density in the Dirac sector is $n_D = 4B_a/\phi_0 = 0.22 \times 10^{12}$ cm$^{-2}$, where $\phi_0 = h/e$ is the flux quantum. The density in the FBs is then $n_F = n_0 - n_D = 3.81 \times 10^{12}$ cm$^{-2}$, which we assign to be the filling factor $\nu = 4$ in the FB. This allows us to determine the twist angle of the tTLG device, $\theta = 1.3°$ (Methods), slightly below the magic angle.

As the Dirac and FB sectors are populated in parallel, the total Chern number is $C = C_D + C_F$, the sum of the Dirac and FB Chern numbers. The $R_{xx} = 0$ line that emanates from the blue circle in Fig. 1c and extends to higher $B_a$ (blue dashed line), has a slope of $C = 2$. At these high $B_a$, only the $N_D = 0$ Dirac LL is occupied with $C_D = 2$, and thus at $\nu = 4$ we find that $C_F = 0$ (as expected due to time reversal symmetry). Additionally, the most prominent Landau fan line emanating from CNP at intermediate $B_a$ is $C = 6$, unlike $C = 4$ in MATBG [1], corresponding to the sum of $C_D = 2$ and $C_F = 4$, further confirming the presence of the 0th LL in the Dirac cone. This is seen further in Extended Data Fig. 1c, which shows line cuts of $R_{xx}$ and $R_{yx}$ at $B_a = 3.2$ T.

Inserting $B_a = 2.25$ T and $N_D = 1$ (Fig. 1c, blue circle) into Eq. 1, we find $\varepsilon_F = W = 49.4$ meV, where $W$ is the bandwidth of the conduction FB. This value is significantly larger than the bandwidth calculated from the noninteracting continuum model of $W \cong 24$ meV (Fig. 2h). This provides a strong indication that Coulomb interactions are at place, as discussed further below. Note that a prominent electron-hole asymmetry is evident in the transport data [11,12]. On the hole side, no pronounced features are observed within the FB, whereas on the electron side a distinct dip in $R_{xx}$ is observed above $\nu = 2$, along with nontrivial $R_{yx}$ (Fig. 1d). Notably, the Dirac LLs change their slope as they cross the $\nu = 2$ peak in $R_{xx}$ (black dotted lines in Fig. 1c), indicating strongly correlated behavior in the FBs [11,12,17], as elaborated below.

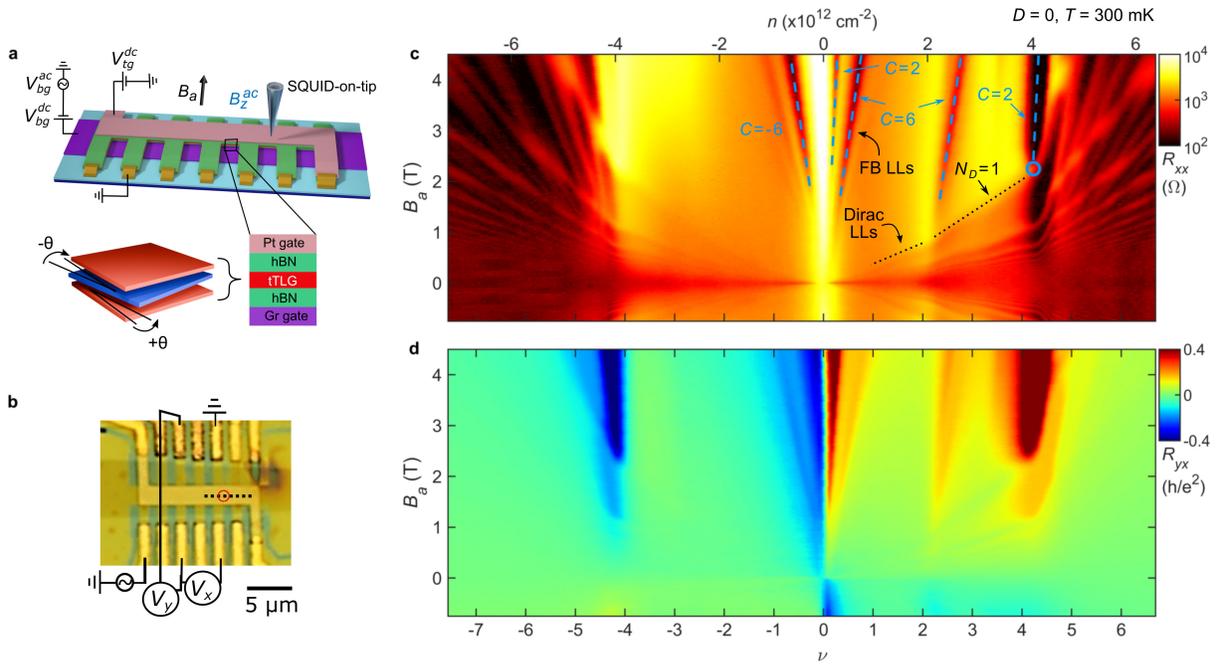

**Fig. 1. Transport measurements in twisted trilayer graphene. a,** Top: Schematic of the tTLG sample. Top gate and bottom gate voltages, $V_{tg}^{dc}$ and $V_{bg}^{dc} + V_{bg}^{ac}$, and the corresponding $ac$ magnetic field $B_z^{ac}$ imaged by the scanning SOT are indicated. Bottom: Schematic of alternating twisted trilayer graphene. **b,** Optical image of the tTLG sample with indicated contacts for $R_{xx}$ and $R_{yx}$ measurements. The dotted black line marks the line cuts in Figs. 2 and 3, and the red circle marks the point measurements in Fig. 4. **c,** $R_{xx}$ vs. carrier density $n$ (top axis) and magnetic field $B_a$ at $T = 300$ mK shown on a logarithmic scale. Two Landau fans emanate from CNP; a steep fan origination from FB LLs and a shallow fan originating from Dirac LLs (seen also Extended Data Fig. 1a). Dashed blue lines follow $R_{xx}$ minima labeled by their Chern number $C$, defined by the slope. The blue circle marks where the $N_D = 1$ Dirac LL crosses the top of the FB. **d,** $R_{yx}$ vs. $\nu$ and $B_a$ in units of $h/e^2$.

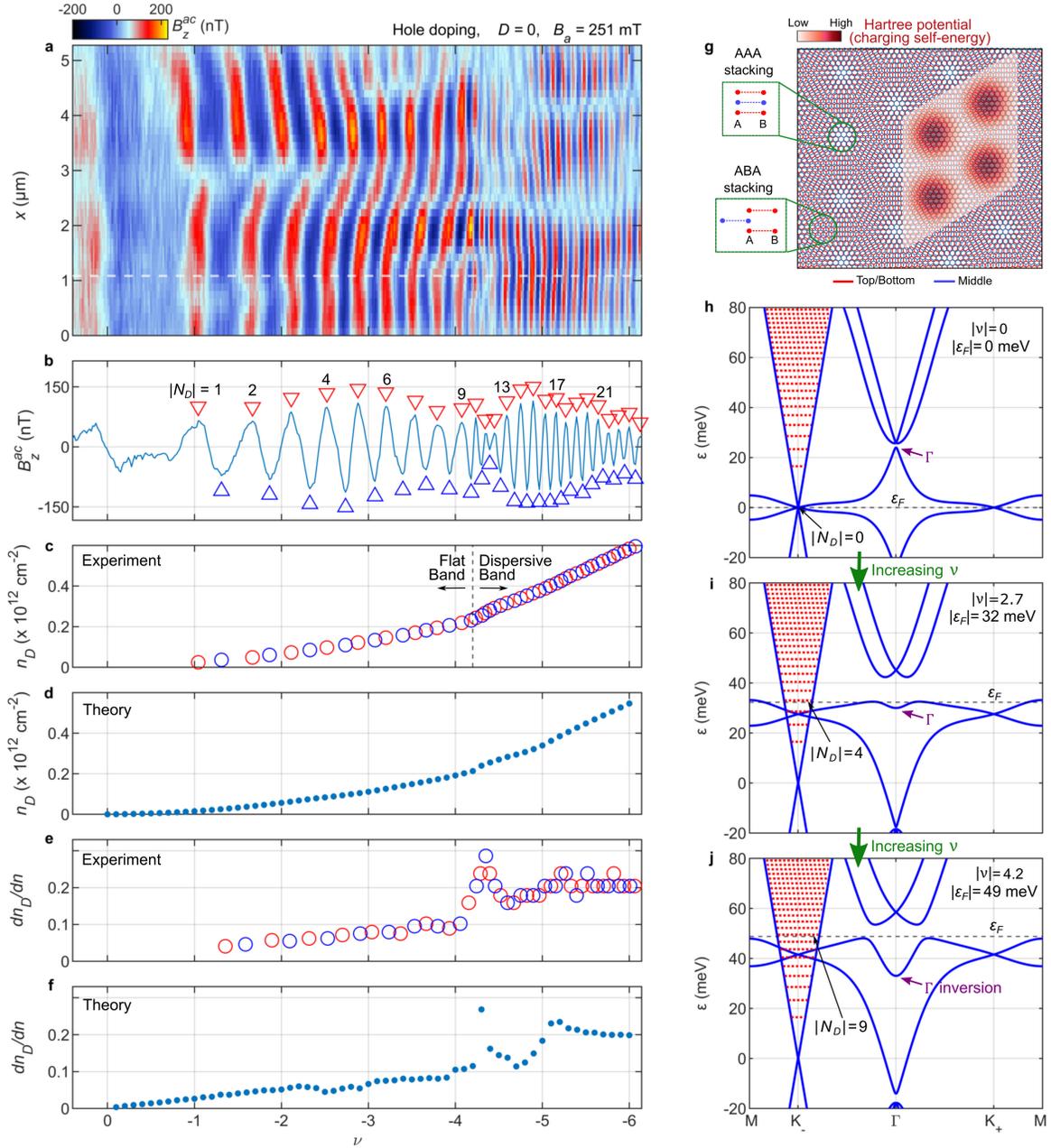

**Fig. 2. Imaging Dirac LLs and Hartree interactions. a,** $B_z^{ac}(x)$ vs. hole carrier density measured along the dotted black line in Fig. 1c at $D = 0$ and $B_a = 251$ mT, reflecting QOs due to LLs in the Dirac band. **b,** Cross-section of $B_z^{ac}$ vs. density along the white dashed line in (a). Red (blue) triangles mark maxima (minima) in the local magnetization due to equilibrium currents flowing in the compressible (incompressible) states in Dirac LLs, labeled by their $|N_D|$. **c,** Carrier density in the Dirac sector, $n_D$, derived from (b). Each oscillation in $B_z^{ac}$ corresponds to adding $\Delta n_D = 4B_a/\phi_0$ carriers to the Dirac sector. Red (blue) circles correspond to $n_D$ extracted from maxima (minima) points in (b). **d,** $n_D$ derived from the Hartree calculation, in good agreement with (c). **e,** Numerical derivative of (c), $dn_D/dn$, which reflects the fraction of carriers added to the system that go to the Dirac sector, as a function of density. Red (blue) circles correspond to derivatives between the adjacent maxima (minima) points in (b). **f,** $dn_D/dn$ extracted from theoretical calculation in (d), in good agreement with (e). **g,** Schematic of the moiré pattern due to the twist angle $\theta$ between the middle and top/bottom layers. The AAA and ABA stacked regions are zoomed in. The wavefunctions of the FB electrons are mostly localized to the AAA regions, leading a charging self-energy described by the periodic Hartree potential shown by the heat map, and to a doping dependent band structure (h-j). **h-j,** Band structure line cuts through high symmetry points of tTLG with Hartree interaction at $|\nu| = 0$ (**h**), 2.7 (**i**), and 4.2 (**j**) showing how the BS changes with doping. At $\nu = 0$ the Hartree interaction term is zero and the BS is equivalent to a single particle continuum model calculation. As $\nu$ is increased the Hartree interaction increases approximately linearly with filling (see Movie 1). Purple arrows bring attention to the inversion of the $\Gamma$ point energy from top of the conduction band at small $\nu$ to the bottom of the conduction FB at large $\nu$. The red dotted lines indicate the LL energy levels $\varepsilon_N^D$ in the Dirac sector at $B_a = 251$ mT. The black dashed line shows the evolution of $|\varepsilon_F|$ with doping and the corresponding Dirac LL $|N_D|$ that it crosses.

**Imaging QOs and the Dirac Landau levels**

In small applied fields, the dense LLs in the Dirac band, described by Eq. 1, can serve as a powerful built-in ruler and spectrometer to probe the underlying physics in the FBs. However, at very low $B_a$ the Dirac LLs are not resolved in transport. To this end, we utilize a scanning superconducting quantum interference device fabricated on the apex of a sharp pipette (SQUID-on-tip, SOT) [52], to detect the dHvA QOs associated with Dirac LLs at $B_a \lesssim 250$ mT. This approach, in addition to offering a fine energy scale, also provides position dependent spectroscopy on the nanoscale [43]. An Indium SOT [53] of about 160 nm diameter is scanned at a height of $h \approx 180$ nm above the sample surface (Fig. 1a) at $T = 300$ mK (Methods). In addition to the *dc* voltages applied to the gates, a small *ac* voltage $V_{bg}^{ac}$ of 85 mV rms is applied to the bottom gate, which modulates the carrier density by $n^{ac}$ corresponding to $\nu^{ac} = 0.03$ rms, and the resulting $B_z^{ac}(x, y) = n^{ac}(dB_z/dn)$ is imaged across the sample.

We explore first the valence FB upon hole doping, where a simpler picture is expected since no symmetry breaking is observed in transport. We measure the evolution of $B_z^{ac}(x)$ with carrier density at $D = 0$ (Fig. 2a) by repeated scanning along the black dotted line in Fig. 1b, while incrementing $\nu$ in 0.02 steps at $B_a = 251$ mT. The oscillations in $B_z^{ac}$ reflect the dHvA QOs in the magnetization due to alternating diamagnetic and paramagnetic equilibrium currents in the compressible and incompressible LL states [54]. Due to the high DOS in the FB sector, and the low $B_a$, only oscillations from the Dirac LLs with larger energy gaps are discerned [44,54]. As $|\nu|$ is increased, the frequency of the $B_z^{ac}(x)$ oscillations increases, and at $|\nu| \simeq 4.2$ there is an abrupt jump in the frequency, as the Fermi energy $\varepsilon_F$ exits the FB and enters the dispersive bands. The spatial variation of the oscillations reflects the inherent disorder in the sample.

Figure 2b shows a cross-section of $B_z^{ac}$ as a function of doping along the dashed white line in Fig 2a. Each maximum (red triangle) marks the center of the $N_D$ compressible LL while each minimum (blue triangle) marks the incompressible state above the same LL. In the range of $|\nu| \lesssim 4.2$, where the Dirac band coexists with the FB, we count nine maxima corresponding to $|N_D| = 9$. Using the LLs in the Dirac band as an energy ruler (Eq. 1), the hole FB thus becomes fully occupied when $|\varepsilon_F|$ reaches $\varepsilon_9^D = 49.5$ meV. This large derived bandwidth $W$ of the FB is in good agreement with transport data and is much larger than the calculated single particle $W \cong 24$ meV, emphasizing the key role of interactions. Similar behavior can be seen in samples 2 and 3 in Extended Data Fig. 2.

Each QO period reflects the filling of an additional Dirac LL, and thus an increase in the Dirac sector carrier density $n_D$ by $\Delta n_D = 4B_a/\phi_0 = 2.4 \times 10^{10}$ cm$^{-2}$. Therefore, by counting the periods upon increasing the total carrier density $n$, we can directly determine $n_D$ as a function of doping, as shown in Fig. 2c. In the FB region ($|\nu| \lesssim 4.2$), $n_D$ grows monotonically with filling, reaching a value of $n_D \cong 0.22 \times 10^{12}$ cm$^{-2}$, in agreement with the value extracted from transport. Upon further doping of the dispersive bands ($|\nu| \gtrsim 4.2$) there is a sharp increase in the slope. This is seen more clearly by plotting the derivative $dn_D/dn$ in Fig. 2e, which can be understood as the fraction of electron that populate the Dirac sector as carriers are added to the system. Such large values of $n_D$ and $dn_D/dn$ cannot be accounted for in a single particle picture and therefore interactions must be considered.

**Charging self-energy and doping dependent BS**

To account for interactions, we begin by considering the charging self-energy of an electron due to charge density already present in the system, which is modeled by a self-consistent calculation of the Hartree term (Methods). A schematic of the moiré pattern (Fig. 2g) shows that the unit cell is comprised of AAA and ABA stacked regions. As the wavefunctions of the FB carriers are mostly localized to the AAA regions, an electron added to the system sees the charge build up on the AAA regions as a background periodic potential [46,55–57]. The strength of the periodic Hartree potential, $V_H$ (Eq. 11 in Methods), is proportional to the charge build up, i.e. carrier density. This interaction driven periodic potential leads to a doping-dependent BS as shown in Figs. 2h-j (see Movie 1 and Methods for BS calculations and parameters). Since the continuum model is electron-hole symmetric, we refer to the conduction band for convenience. At the CNP the charging self-energy is zero, and electronic states at the $\Gamma$ point are at the highest energy in the flat conduction band (Fig. 2h, purple arrow). As carrier density is increased, the charging self-energy increases and states away from the $\Gamma$ point, which are mostly localized to the AAA regions, have their energies increased due to electronic buildup on the AAA sites, eventually causing an inversion of the $\Gamma$ point to reside at the bottom of the flat conduction band (Fig. 2j, purple arrow). This effect is accompanied by a large overall increase of the bandwidth of the FB relative to the original single particle $W$ in the absence of interactions (Fig. 2h) [15,57].

From the Hartree calculation we extract the evolution of $\varepsilon_F$ with filling, which crosses the top of the flat conduction band at an energy ~49 meV consistent with the bandwidth $W$ extracted from experiment. As the Dirac cone is static, $\varepsilon_F$ directly translates into the Dirac cone density, $n_D = \frac{4S_D}{4\pi^2} = \pi k_F^2/\pi^2 = \frac{1}{\pi}\left(\frac{\varepsilon_F}{\hbar v_F}\right)^2$, where $S_D$ is the $k$ space area of the Dirac cone, $k_F$ is the Fermi crystal momentum, and the factor 4 comes from the four-fold degeneracy of the Dirac cone. $n_D$ vs. doping is plotted in Fig. 2d, which matches experiment well (Fig. 2c). Furthermore, Fig. 2f shows the derivative $dn_D/dn$, which shows the fraction of the carriers that populate the Dirac band upon additional doping. $dn_D/dn$ increases as function of carrier density, reaching values of 0.1 for $|\nu| \lesssim 4.2$. In a non-interacting system, the BS is static and $dn_D/dn$ would be given by the ratio of the Dirac DOS to the total DOS, and be at most about 0.005 due to the extremely high DOS of the FBs, which causes $\varepsilon_F$ to change slowly with doping. When accounting for the charging self-energy, however, the DOS of the FBs plays only a minor role since the BS evolves with doping. The charging energy causes $\varepsilon_F$ to increase with carrier density at a much faster rate, as the FBs are pushed up in energy, transferring more carriers into the Dirac sector (Fig. 2h-j and Movie 1). For $|\nu| \gtrsim 4.2$ the jump in $dn_D/dn$ is well reproduced by the theory in Fig. 2f, reflecting the drop in the total DOS at the Fermi level upon entering the dispersive bands. The agreement between the highly sensitive theoretical and experimental $dn_D/dn$ for such an extended range in $\nu$, through the FBs and into the dispersive bands, is striking. This emphasizes the importance of the charging self-energy in providing a complete description of the mirror symmetric twisted graphene family, which shows the strongest superconductivity of all moiré systems [21].

**Symmetry breaking at half filling**

We now explore the electron doped spectrum, which shows signatures of symmetry breaking in transport. Figure 3a shows $B_z^{ac}(x)$ vs. electron doping acquired at a lower $B_a = 131$ mT, which provides higher energy resolution utilizing denser Dirac LLs (see Extended Data Fig. 3 for $B_a = 251$ mT data). For $\nu \lesssim 2$ the oscillations in $B_z^{ac}$ are similar to their hole counterpart, while for $2 \lesssim \nu \lesssim 2.6$ a bunching of QOs is observed, along with increased intensity. This is seen more clearly in the cross-section of $B_z^{ac}$ as a function of doping (Fig. 3b) along the dashed white line in Fig. 3a, and in the derived $dn_D/dn$ (Fig. 3c). For both low ($\nu \lesssim 2$) and high fillings ($\nu \gtrsim 2.6$) $dn_D/dn$ follows a behavior similar to the valence band (Fig. 2e). However, at half filling of the conduction FB there is a clear jump in $dn_D/dn$ (shaded region), which is absent in the valence band (Fig. 2e), which means that a larger fraction of carriers added to the system populate the Dirac sector. This is a clear signature of interaction-driven degeneracy lifting.

Following the data, we take as an ansatz a simple Stoner instability model where for $2 \leq \nu \leq 2.6$ one flavor is increased in energy by $\Delta_{HF}$ and the other is decreased by the same (Figs. 3e-g and Methods). Based on HF calculations at even integer fillings in magic angle twisted systems [20,23,25,27,30], we take $\Delta_{HF} = 20$ meV, which we add to the Hartree calculated bands (Fig. 3f). The resulting occupation of the two flavors in this simple Stoner model is sketched in Fig. 3g, where the shaded region marks the symmetry broken state. Figure 3f shows that in the symmetry broken state $\varepsilon_F$ is much larger than in the symmetric state (Fig. 3e), and consequently more carriers are transferred into the Dirac sector. This results in a clear increase in the calculated $dn_D/dn$ shown in Fig. 3d (shaded region), which is in good agreement with the data in Fig. 3c. The enhancement of $dn_D/dn$ is caused by the aforementioned upward shift in $\varepsilon_F$ in the symmetry broken state. Additionally, this mechanism is consistent with the increased amplitude of the $B_z^{ac}$ oscillations, as in the symmetry broken state there are fewer states in the FB sector to tunnel to, resulting in lower Dirac LL broadening and larger signal.

**Displacement field dependence**

Next, we explore the BS evolution upon applying a displacement field $D$. We station the SOT at a fixed position (red circle, Fig. 1b), and measure $B_z^{ac}$ oscillations as a function of $\nu$ and $D$ at $B_a = 251$ mT (Fig. 4a) and at a lower $B_a = 56$ mT for better energy resolution with much denser LLs (Fig. 4b). We begin by observing that at $B_a = 251$ mT and $D \approx 0$ (dashed box), the QOs as a function of $\nu$ are similar to those in Figs. 2a and 3a, as expected. However, the $B_a = 56$ mT data at $D \approx 0$ (dashed box) does not scale as expected. From Eq. 1 it follows that the $N^D = 1$ LL at 56 mT should appear at a lower energy by a factor of the square root of the ratio of $B_a$, i.e. $\sqrt{251/56} \approx 2.1$. In the case of a high and uniform FB DOS, the $N^D = 1$ LL that appears at $\nu = 0.95$ at $D \approx 0$ in Fig. 4a, should accordingly appear at $\nu = 0.45$ at 56 mT. Strikingly, Fig. 4b shows that the $N^D = 1$ LL appears at $\nu = 0.27$, almost a factor of 2 below the above estimate. This strongly indicates that considering only the charging self-energy, which was used to elucidate our results thus far, cannot fully describe the correlated physics in this system close to CNP.

Now we move to analyze the dispersion of the Dirac LLs with increasing $|D|$, or equivalently, as a function of outer layer potential difference $U$. Finite $D$ breaks the mirror symmetry and hybridizes the Dirac and FB sectors. In models where the graphene and FB Dirac cones overlap at the mini Brillouin zone (mBz) corners ($K_+, K_-$), such as single particle and Hartree only calculations, the hybridization occurs at the CNP (Extended Data Fig. 4). With increasing $U$ the Dirac node gets pushed to higher energy. As a result, the $0^{th}$ Dirac LL along with all the higher Dirac LLs are pushed away from CNP (Fig. 4c, grey dashed), similar to ABA trilayer graphene [44].

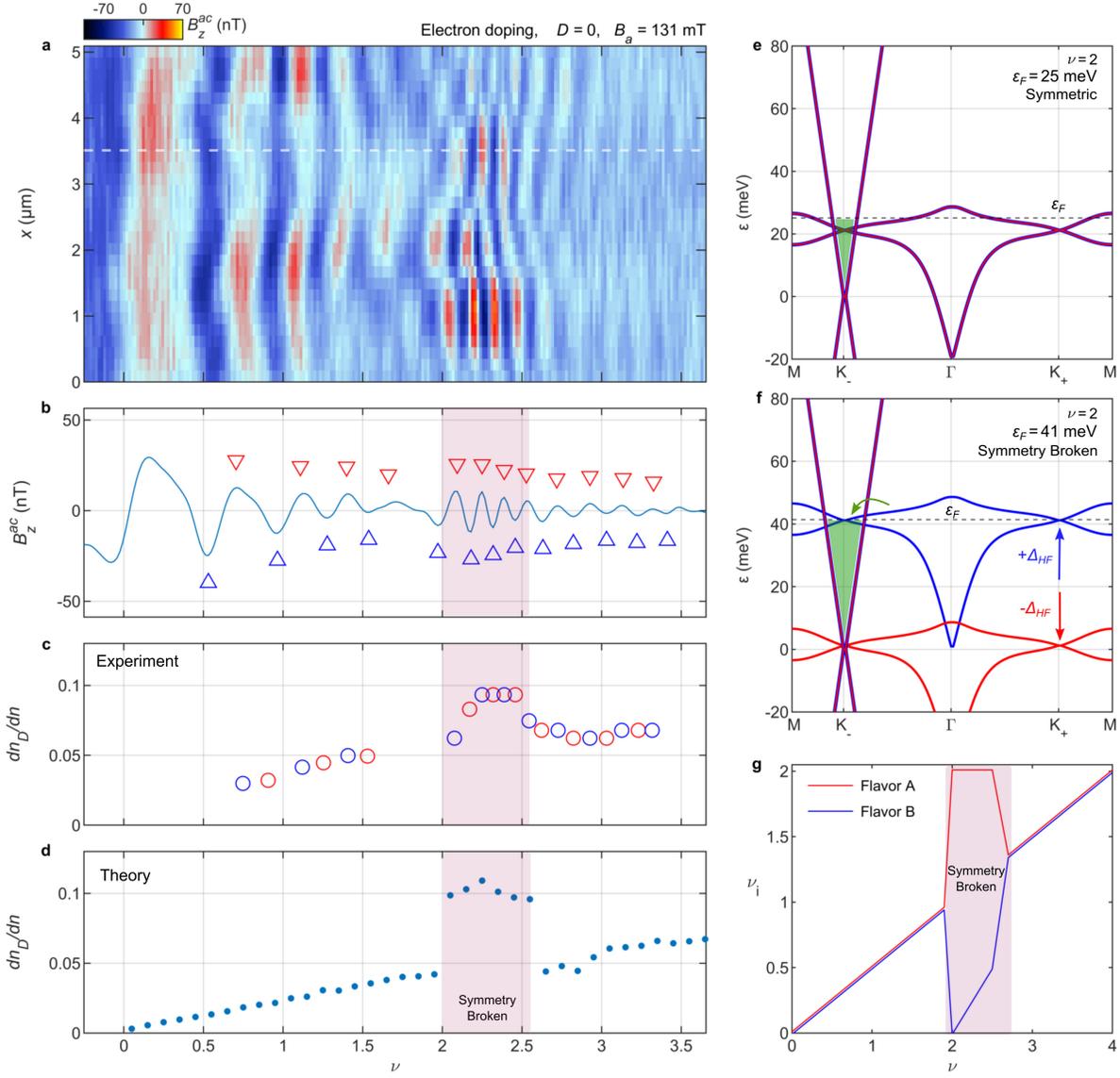

**Fig. 3. Symmetry breaking at $\nu = 2$. a,** $B_z^{ac}(x, \nu)$ QOs like in Fig. 2, but for electron doping and at a lower $B_a = 131$ mT, which provides higher resolution for analyzing the symmetry breaking at $\nu = 2$. **b,** Cross-section of $B_z^{ac}(\nu)$ along the white dashed line in (a) with red (blue) triangles marking local maxima (minima). A Gaussian filter is used to smooth the data and locate the extrema. **c,** The fraction of electrons populating the Dirac sector as electrons are added to the system, $dn_D/dn$, extracted from maxima (red) and minima (blue) points in (b). **d,** $dn_D/dn$ derived from the Hartree calculation, assuming an ansatz of a $\pm 20$ meV flavor degeneracy lifting for $2 \leq \nu \leq 2.6$ (shaded region). **e-f,** Band structure at $\nu = 2$ with the proposed Stoner ansatz. Compared to the symmetric case (**e**), in the symmetry broken state (**f**) the energy of one flavor (A) is decreased by 20 meV while the other flavor (B) is increased by 20 meV. In the symmetry broken state $\varepsilon_F$ increases and hence additional carriers are transferred into the Dirac cone (shaded green). **g,** Schematic of the occupation $\nu_i$ of the different flavors with the ansatz of symmetry breaking for $2 \leq \nu \leq 2.6$ (Methods).

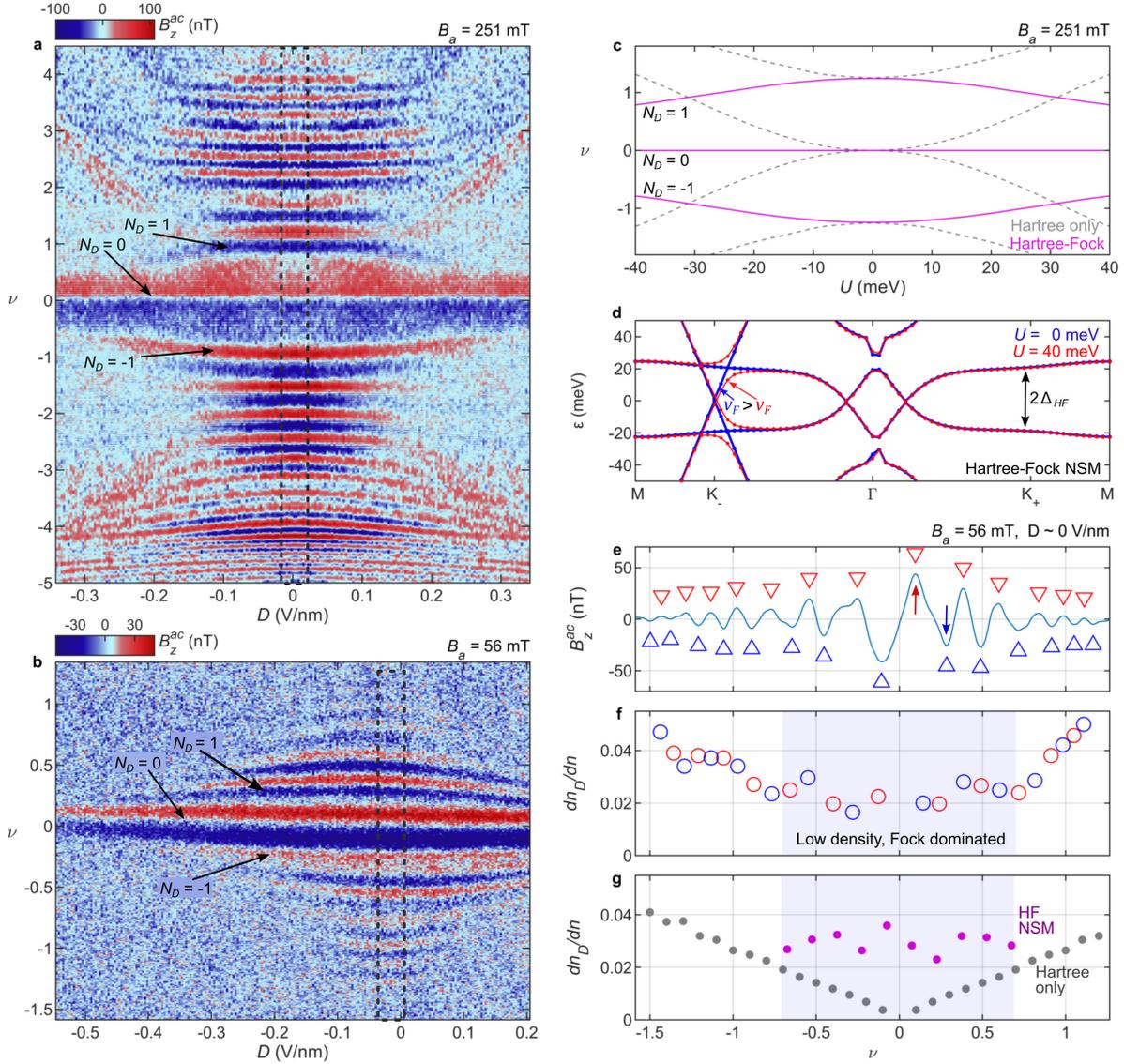

**Fig. 4. Displacement field dependence and Hartree-Fock calculations. a,** $B_z^{ac}$ QOs due to LLs in the Dirac band measured at a fixed SOT position (red circle, Fig. 1b) as a function of $\nu$ and $D$ at $B_a = 251$ mT. Diagonal streaks are an artifact coming from the bottom gate. The $0^{th}$ Dirac LL has no $D$ dependence, the $N_D = \pm 1$ LLs curves toward CNP, while higher LLs curve away from CNP. **b,** Same as (a) but at $B_a = 56$ mT. The $0^{th}$ LL has no $D$ dependence, but all other LLs curve towards CNP. **c,** Simulation of Dirac LLs as a function of the layer potential difference $U$ at $B_a = 251$ mT, using the NSM state obtained from a full HF treatment (purple) and the Hartree interaction only (dashed grey). The evolution of $N_D = 0, \pm 1$ LLs in experiment is consistent with the simulation for NSM. **d,** Band structure of NSM state (Extended Data Fig. 5b) at $U = 0$ (blue) and $U = 40$ meV (red). Flat band Dirac cones migrate towards $\Gamma$ and the HF gap at $K_+, K_- \approx 2\Delta_{HF} = 40$ meV. $U$ induced hybridization between the Dirac cone and FBs (red) causes a reduced $v_F$ in the Dirac cone below the hybridization energy (Extended Data Fig. 5a). **e,** Cross-section of $B_z^{ac}$ averaged over a range around $D \approx 0$ in the dashed box in (b). The red arrow marks the first electron doped paramagnetic peak and the blue arrow marks the diamagnetic peak at $N_D = 1$. **f,** $dn_D/dn$ extracted from (e). Shaded region marks low

carrier density where the Hartree interaction is negligible and the Fock Interaction dominates. **g,** Theoretical calculation of $dn_D/dn$ using Hartree interaction BS (grey) and the Fock induced NSM BS (purple).

Figure 4a indeed shows that in experiment the higher LLs disperse away from CNP as expected. However, there are two peculiarities. First, the $0^{th}$ LL stays pinned at CNP independent of $D$. Second, instead of curving away from CNP, the $N_D = \pm 1$ LLs have the opposite behavior, curving towards the CNP.

At lower $B_a = 56$ mT (Fig. 4b), we confirm that the $N_D = 0$ LL does not disperse, and in addition, all the detectable $|N_D| > 0$ LLs curve towards the CNP. Combining the information from Figs. 4a,b, we conclude that the Dirac LLs residing at $|\nu| \lesssim 1.5$ curve towards CNP, whereas LLs appearing at $|\nu| \gtrsim 1.5$ curve away from the CNP, in sharp contrast to the Hartree prediction in Fig. 4c (grey dashed). The higher carrier density behavior is thus apparently captured well by considering only charging self-energy, as done in Figs. 2,3. However, as the charging self-energy becomes small near CNP, it is necessary to consider exchange interactions to gain insight into the underlying physics.

**Exchange interactions**

The observed displacement field dependence reveals lack of hybridization at CNP, which means that graphene and FB Dirac cones do not overlap at zero energy. This can arise due to either gapping the FB Dirac cones, namely forming an insulating FB state, or moving them away from the mBz corners ($K_-, K_+$), maintaining a semimetallic FB phase. Note that the latter automatically breaks $C_3$ rotational symmetry independent of the exact nature of the ground state. To consider possible ground states, we point out that at CNP the charging self-energy is zero but strong exchange interactions in the FBs, modeled by the mean field Fock interaction term, may drive symmetry broken orders. HF calculations at the CNP in tTLG allow many candidate symmetry-broken ground states, including valley polarization (VP), valley Hall (VH), Kramer intervalley coherence (KIVC), and the nematic semimetal (NSM) [20]. We first analyze the NSM ground state, in which exchange energy causes the FB Dirac cones to migrate toward the $\Gamma$ point (Extended Data Fig. 5b), spontaneously breaking $C_3$, and no longer overlapping the graphene Dirac cones. As a result, at $U = 0$ the Dirac and FBs intersect at an energy equal to $\Delta_{HF} \cong \pm 20$ meV (Fig. 4d (blue), Extended Data Fig. 5d). Consequently, at $U \neq 0$ the hybridization occurs at energies around $\Delta_{HF}$ (Fig. 4d, red curve), rather than near the graphene Dirac nodes as in the Hartree only or single particle cases (Extended Data Fig. 4). As $U$ is increased, the hybridization strength increases, effectively lowering $v_F$ (or equivalently the slope) of the Dirac cone below the hybridization energy, as can be seen in Fig. 4d (red).

This picture successfully explains the low energy experimental observations. First, the graphene Dirac node never hybridizes, so the $N_D = 0$ LL does not disperse. Second, increasing $|D|$ decreases $v_F$ below $\Delta_{HF}$, thereby reducing the energy of the Dirac LLs (Eq. 1). This can be seen in Fig. 4c (purple), which shows a simulation of the low energy Dirac LLs at $B_a = 251$ mT, by calculating the renormalized $v_F$ as a function of $|D|$ in the NSM state (Extended Data Fig. 5a). Finally, the experimentally observed boundary of $|\nu| \cong 1.5$ separating the upward and downward dispersing Dirac LLs corresponds to a hybridization energy of about 19 meV, in good agreement with $\Delta_{HF}$ from simulation.

Note that other HF ground states similarly result in a renormalization of $v_F$ with $U$. These states, such as KIVC, have gapped FB Dirac cones, also leading to hybridization effect at higher energy $\Delta_{HF}$ (Extended Data Fig. 5c). Therefore, the $D$ field dependence measurements provide irrefutable evidence that the FB Dirac cones either break $C_3$ and migrate away from the mBz corners, as in the NSM case, or have been gapped, as in the KIVC case.

**Ground state at CNP**

We return focus to the $D \approx 0$ segment of Fig. 4b, with Fig. 4e showing a cross-section of $B_z^{ac}$ averaged around $D \approx 0$ in the dashed black box. The dense LLs at $B_a = 56$ mT carry a wealth of information about the correct ground state. Note that at this $B_a$, the $N_D = 1$ LL resides at energy $\varepsilon_1^D =$7.8 meV, observed as diamagnetic peak in the QOs (blue arrow). Furthermore, the first paramagnetic peak, due to equilibrium currents flowing in the incompressible gap between the $N_D = 0$ and 1 LLs (red arrow), can be approximated to be at energy $\varepsilon_1^D/2 = 3.9$ meV. Figure 4e shows that this paramagnetic peak appears at density $\nu \cong 0.1$ corresponding to $n = 9 \times 10^{10}$ cm$^{-2}$, while only $n_D = 2B_a/\phi_0 = 2.7 \times 10^9$ cm$^{-2}$ carriers reside in the Dirac band. This means that at this energy most of the carriers reside in the FB, namely the FB has a significant DOS below 3.9 meV. Hence, if a gap exists it must be less than a few meV. In fact, a more careful detailed analysis of the QO data allows us to place an even lower bound of <1 meV on the possible gap at CNP (see Methods). This bound seems incompatible with a correlated gapped phase, where the gap is expected to be comparable to the Coulomb repulsion energy scale ~20 meV [23,30].

Furthermore, in our device, we observe a separation between the upward and downward dispersing Dirac LLs with $D$ at a boundary of $|\nu| \cong 1.5$, which corresponds to a hybridization energy of about 19 meV. This means that the FBs have a $K$-point gap of about ±19 meV. There is no generic theoretical reason for an order of magnitude difference between the $K$-point gap and the $\Gamma$-point gap in the FBs. In the vast theoretical literature to date concerning mirror-symmetric twisted graphene, there exist no ground state candidates with K-point gap ~19 meV and a tiny global gap. These theoretical and experimental considerations provide strong evidence for the lack of a gap in the FBs, and effectively rules out the KIVC or any other gapped ground state at CNP. Following our above findings, we thus conclude that the CNP ground state is a $C_3$ breaking NSM.

To further verify this conclusion, we analyze the full range of QOs near the CNP at $B_a = 56$ mT. Similar to Figs. 2 and 3, we extract experimentally $dn_D/dn$ (Fig. 4f) for the entire set of oscillations in Fig. 4e. We focus on the lowest Dirac LLs $|N_D| \leq 3$. At such low densities, we approximate the charging self-energy to be zero and view the QOs as a reflection of a static BS renormalized by the exchange energy. Here, $dn_D/dn$ is simply the relative DOS of the Dirac sector in the static BS which can be readily calculated from simulation. Figure 4g shows $dn_D/dn$ calculated for the Hartree term only (grey), and for HF NSM (purple). For large $\nu$ ($|N_D| > 3$), the charging self-energy becomes dominant and the data follows $dn_D/dn$ calculated from Hartree term only. Close to CNP, however, the HF NSM has the correct FB DOS to qualitatively match experiment. This provides further corroboration that the ground state at CNP is indeed NSM.

**Discussion**

Generally, HF analytics find the KIVC or VH state to be energetically favorable with respect to the NSM state [23]. However, even small amounts of heterostrain drive a transition that stabilizes the NSM [51]. Twisted systems in general have been shown to have significant strain [58], including tTLG where STM measurements show that angle mismatch often relaxes into the mirror symmetric configuration [14]. The apparent strain is likely related to the spatial variations in the QOs observed in Figs. 2a and 3a. Our finding of a NSM ground state at CNP is consistent with these considerations, and consistent with most experiments that find no evidence of a gap at CNP [2,4,7,8].

The regime of slightly off magic-angle graphene, where correlated-electron effects are weaker, provides opportunities to study symmetry breaking instabilities, as well as their hierarchy. Our finding that the Stoner transition persists even in the absence of insulating states (Fig. 3), establishes the Stoner polarized states as the parent state for the emergence of correlated insulators. By reducing the interaction strength, we find the transition near half filling to be most robust, in agreement with the appearance of the strongest correlated insulator state in MATBG at the same filling.

The high sensitivity of the thermodynamic QOs to the band structure makes our measurement technique a powerful probe of interaction effects in highly correlated systems. The remarkable detection of local QOs at $B_a$ as low as 56 mT allows us to explore questions surrounding symmetry breaking and fragile ground states in graphene systems that have yet remained unsolved. Our findings lead to a framework that isolates the different interaction effects, where high carrier doping regime can be well described by considering the charging self-energy, whereas near CNP the exchange energy becomes dominant.

This technique can readily be generalized to interacting systems that do not naturally contain a Dirac cone in the BS. One can add a Dirac band to essentially any vdW system by placing an additional monolayer graphene sheet twisted at a large angle [59,60], such that the graphene and the system of interest share charge density but are effectively isolated from each other at low energy. The potential of low-field QOs in providing invaluable information about the BS and interactions, calls for the development of new methods with similarly high sensitivity.

**Acknowledgments** We thank Peleg Emanuel for fruitful discussions. This work was co-funded by the Minerva Foundation grant No 140687, by the United States - Israel Binational Science Foundation (BSF) grant No 2022013, and by the European Union (ERC, MoireMultiProbe - 101089714). Views and opinions expressed are however those of the author(s) only and do not necessarily reflect those of the European Union or the European Research Council. Neither the European Union nor the granting authority can be held responsible for them. E.Z. acknowledges the support of the Andre Deloro Prize for Scientific Research, Goldfield Family Charitable Trust, and Leona M. and Harry B. Helmsley Charitable Trust grant #2112-04911. Y.R. acknowledges the funding received from the MINERVA Stiftung with the funds from the BMBF of the Federal Republic of Germany. K.W. and T.T. acknowledge support from the JSPS KAKENHI (Grant Numbers 20H00354, 21H05233 and 23H02052) and World Premier International Research Center Initiative (WPI), MEXT, Japan. GS acknowledges support from the Walter Burke Institute for Theoretical Physics at Caltech and from the Yad Hanadiv Foundation through the Rothschild fellowship. Y.O. acknowledges support from the European Union's Horizon 2020 research and innovation programme (Grant Agreement LEGOTOP No. 788715), and the DFG (CRC/Transregio 183, EI 519/7-1). D.E.P. is supported by the Simons Collaboration on UltraQuantum Matter, which is a grant from the Simons Foundation, and startup funding from the University of California at San Diego. MB acknowledges the VATAT Outstanding PhD Fellowship in Quantum Science and Technology.

**Author contributions** I.R., M.B. and E.Z. designed the experiment. I.R. and M.B. performed the measurements. V.B., M.U. and Y.R. designed and fabricated the sample and contributed to the analysis of the results. N.B and I.R. fabricated the SOTs and Y.M. fabricated the tuning forks. Y.Z performed the single particle calculations. G.S., M.B. and Y.O. performed the Hartree calculations. P.L and D.P. performed the Hartree-Fock calculations. K.W. and T.T. provided the hBN crystals. M.B., I.R., G.S., P.L and D.P. performed the analysis and theoretical modeling. M.B., I.R., G.S., P.L, D.P. and E.Z. wrote the original manuscript. All authors participated in discussions and revisions of the manuscript.

**Competing interests** The authors declare no competing interests.

**Data availability** The data that supports the findings of this study are available from the corresponding authors on reasonable request.

**Code availability** The Hartree potential and HF calculations used in this study are available from the corresponding authors on reasonable request.

**Methods**

**Device fabrication**

The hBN-encapsulated tTLG device was fabricated using the dry-transfer method. The flakes were first exfoliated onto a Si/SiO$_2$ (285 nm) substrate and picked up using a polycarbonate on a polydimethylsiloxane (PDMS) dome stamp. The number of layers in the graphene flakes was determined by Raman microscopy, and the crystallographic orientation of both hBN and tTLG were determined from their straight edges. The WITec alpha300 R Raman Imaging Microscope was used to carry out Raman measurements using wavelengths of 532 nm, and to cut monolayer graphene into three pieces using 1064 nm laser beam. During the dry-transfer process, the crystal axes of each layer of the tTLG stack were aligned by a mechanical rotation stage. After encapsulation with hBN the stacks were released onto a pre-annealed graphite local bottom gate (~7-10 nm) patterned on Si/SiO$_2$ wafer for Devices 1 and 2. For Device 3, p-doped Si was used as the bottom gate. The finalized stacks were annealed in vacuum at 350 °C for strain release. A Ti (2 nm)/Pt (12 nm) top gate was then deposited on top of the stack for Device 1. For Devices 2 and 3, the top gate is Ti (4nm)/Au (14 nm). The 1D contacts were etched by SF$_6$ and O$_2$ plasma, followed by deposition of contacts of Cr (4 nm)/Au (70 nm) in angle rotated e-gun evaporator. Consequently, the Hall bar geometry was etched using SF$_6$ and O$_2$ plasma. Finally, the surface resist and etching residues were swept off by atomic force microscopy (AFM) in contact mode.

Device summary:

**Device 1** (presented in the main text): Graphite bottom gate, Ti/Pt top gate; bottom hBN thickness ≈ 55 nm, bottom gate capacitance $C_{bg} \approx 3.4 \times 10^{11}$ e·cm$^{-2}$·V$^{-1}$, top hBN thickness ≈ 32 nm, top gate capacitance $C_{tg} \approx 5.8 \times 10^{11}$ e·cm$^{-2}$·V$^{-1}$, twist angle $\theta_M \approx 1.3°$. The capacitances were calculated by fitting the slopes of Chern insulator lines in the magnetotransport measurements.

**Device 2**: Graphite bottom gate, Ti/Au top gate; bottom hBN thickness ≈ 36 nm, $C_{bg} \approx 5.2 \times 10^{11}$ e·cm$^{-2}$·V$^{-1}$, top hBN thickness ≈ 60 nm, $C_{tg} \approx 3.1 \times 10^{11}$ e·cm$^{-2}$·V$^{-1}$, twist angle $\theta_M \approx 1.5°$.

**Device 3**: p-doped Si bottom gate, Ti/Au top gate; bottom hBN thickness ≈ 36 nm, $C_{bg} \approx 7.5 \times 10^{10}$ e·cm$^{-2}$·V$^{-1}$, top hBN thickness ≈ 60 nm, $C_{tg} \approx 3.4 \times 10^{11}$ e·cm$^{-2}$·V$^{-1}$, twist angle $\theta_M \approx 1.5°$.

**Transport measurements**

Four-probe transport measurements were performed at $T = 300$ mK using standard lock-in technique with an *ac* bias current of $I^{ac} = 10$ nA rms at ≈ 11 Hz. Both longitudinal and transverse resistivity ($R_{xx}$ and $R_{yx}$) are measured as a function of carrier density $n$ and out-of-plane magnetic field $B_a$. From the $R_{xx}$ measurements we distinguish the Dirac sector from the FBs and find the full filling of the FBs at $n_F = n_0 - n_D = 3.81 \times 10^{12}$ cm$^{-2}$. We assign this density $n_F = 4\nu = 4/A_m$, where $A_m$ is the moiré unit cell area and is connected to the small twist angle $\theta$ by $A_m = \frac{\sqrt{3}a^2}{2\theta^2}$, where $a = 0.246$ nm is the graphene lattice constant. From this we determine the twist angle of the tTLG Device 1 of $\theta = \sqrt{\frac{\sqrt{3}a^2}{2A_m}} \approx 1.3°$, slightly below the magic angle.

The transport measurements in Fig. 1 and $B_z^{ac}$ line scan data in Figs. 2 and 3 were done at $D = 0$, while the point scan data in Fig. 4 of $B_z^{ac}$ are as a function of $\nu$ and $D$. The displacement field is defined in terms of the top gate voltage, $V_{tg}^{dc}$, and bottom gate voltage, $V_{bg}^{dc}$, as $D = \left( C_{tg}(V_{tg}^{dc} - V_{tg}^0) - C_{bg}(V_{bg}^{dc} - V_{bg}^0) \right)/$ $2\epsilon_0$. Here $\epsilon_0$ is the permittivity of free space, and $V_{tg}^0$ and $V_{bg}^0$ correspond to the charge neutrality of the top and bottom gates respectively obtained from transport measurements at zero magnetic field.

Extended Data Fig. 1a shows $dR_{xx}/dB_a$ of the $R_{xx}$ data shown in Fig. 1c. The derivative data shows clearly the LL fan emanating from the Dirac sector. Similar to Ref. [11] a peak in $dR_{xx}/dB_a$ appears when a compressible Dirac LL passes through a compressible FB state, whereas a dipole-like feature is observed when the FB is gapped (near $\nu = 4$), as well as in the remote bands. The yellow circle marks where the $N_D = 1$ LL crosses the top of the FB, which is used to extract the bandwidth of the FB. A clear kink in the Dirac LLs can be seen upon passing through the $\nu = 2$ symmetry broken state.

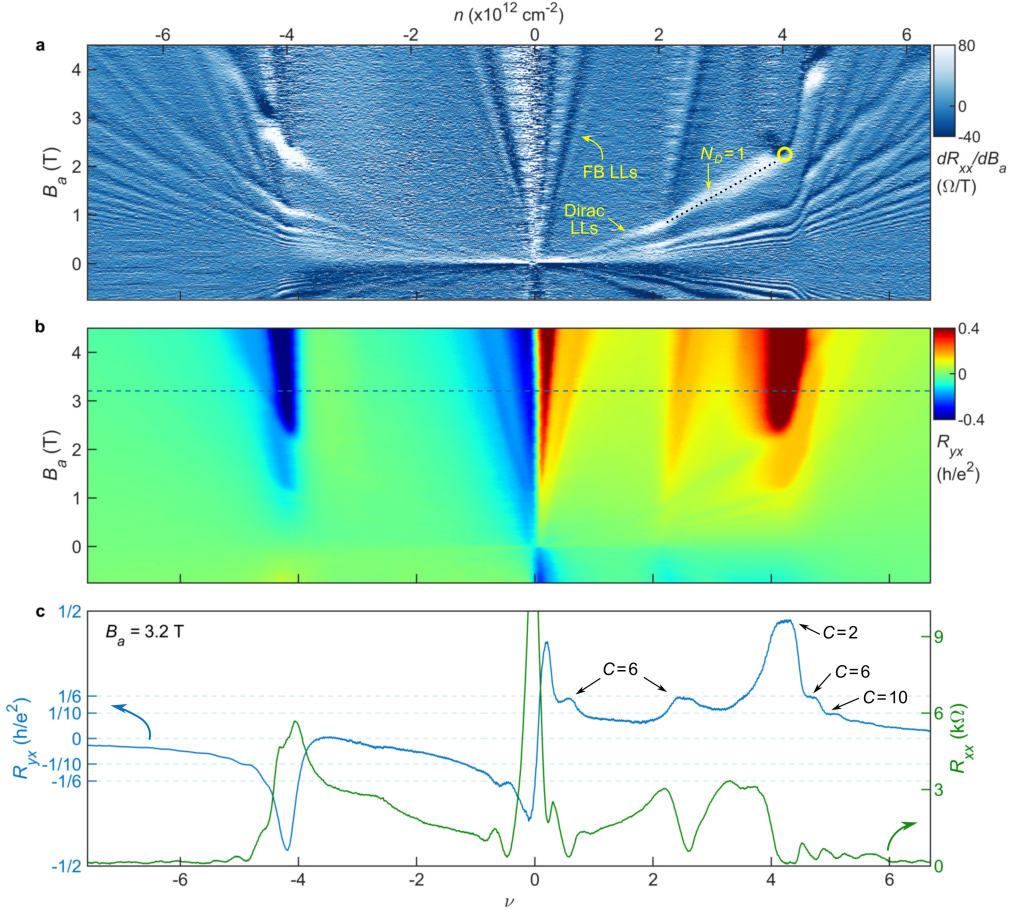

**Extended Data Fig. 1. Transport and Chern numbers. a,** Numerical derivative of $R_{xx}$ shown in Fig. 1c, $dR_{xx}/dB_a$. Here the Dirac LLs are readily discerned. The yellow circle marks the crossing of the $N_D = 1$ Dirac LL with the top of the FB. **b,** $R_{yx}$ reproduced from Fig. 1d. **c,** $R_{xx}$ (right axis) and $R_{yx}$ (left axis) vs. $\nu$ at $B_a = 3.2$ T along the dashed lines in (b). Quantized $R_{yx}$ values are indicated with their Chern numbers that match the accompanying slopes of the $R_{xx}$ minima in Fig 1c.

Extended Data Fig. 1b shows a replica of the $R_{yx}$ data from Fig. 1, along with linecuts of both $R_{xx}$ and $R_{xy}$ at $B_a = 3.2$ T (Extended Data Fig. 1c), where only the $N_D = 1$ Dirac LL is filled. Above the full filling of the FB ($\nu \gtrsim 4.2$) a sequence of quantized $R_{yx}$ and dips in $R_{xx}$ is observed with $C = 2, 6, 10$, corresponding to $C_F = 0$ and $C_D = 2, 6, 10$. As discussed above, a prominent electron-hole asymmetry is evident, with no correlated states observed in the valance FB. On the electron side however, at elevated fields $B_a \gtrsim 1.5$ T, and $R_{xx}$ minimum follows a $C = 6$ slope and the corresponding $R_{yx}$ is quantized with the same Chern number (Extended Data Fig. 1c). At these fields, only the $N_D = 0$ LL is filled, giving $C_D = 2$ and $C_F = 4$. A topological band with such a high Chern number $C_F$ is surprising, and might be explained by translation symmetry breaking that doubles the number of low energy bands [61], but is not the focus of this work. Our local QO measurements explore low fields, where $R_{yx}$ quantization disappears and the slope in $R_{xx}$ more closely follows a $C = 2$ line (Extended Data Fig. 1a at $B_a < 1$ T), suggesting a different ground state.

**SOT fabrication and QO measurements**

The $ac$ magnetic field measurements were done using Indium SOT of 160 nm effective diameter, which was fabricated as described previously [52,53,62]. The imaging was performed at $T = 300$ mK using a cryogenic SQUID series array amplifier (SSAA) [63]. The SOT had a magnetic field sensitivity down to 10 nT/Hz$^{1/2}$ and was attached to a quartz tuning fork excited at its resonance frequency of $\approx 33$ kHz for height control as described in [64]. The scanning was performed at a height of 180 nm above the sample surface.

The $B_z^{ac}$ images were obtained with pixel size of about 90 nm and acquisition time of 1 s/pixel. The measured signal $B_z^{ac} = n^{ac}(dB_z/dn)$ is proportional to the modulation in the carrier density $n^{ac}$ induced by a small $ac$ voltage $V_{bg}^{ac}$ of 85 mV rms applied to the backgate at a frequency of $f \approx 5$ kHz. To obtain the best signal to noise ratio from the QOs in the Dirac band, the $V_{bg}^{ac}$ amplitude was optimized for the lowest $B_a = 56$ mT, for which the period of the QOs, $\Delta n$, is the smallest.

**Magnetization oscillations in Device 2 and 3**

Similar to Device 1, thermodynamic QOs are observed in Devices 2 and 3 (Extended Data Fig. 2). Upon hole doping, a monotonic increase in $dn_D/dn$ is observed (Extended Data Figs. 2b,e), similar to Device 1 (Fig. 2). In contrast, electron doping displays jumps in $dn_D/dn$ indicating symmetry breaking. Interestingly, in Devices 2 and 3, $dn_D/dn$ peaks occur both at $\nu = 2$ and 3 (green circles). This matches transport data for Device 3, where at high $B_a$ insulating states appear near $\nu = 1$, 2, and 3. This is likely because the twist angle in these devices ($\theta \approx 1.5°$) is closer to the magic angle.

**Observation of symmetry breaking in Device 1**

Figure 3 shows a peak in $dn_D/dn$ signifying a symmetry breaking at $\nu = 2$ measured at $B_a = 131$ mT. The same behavior is observed at higher $B_a = 251$ mT (Extended Data Fig. 3), albeit with lower resolution due to sparser LLs at higher $B_a$. The precise ground state of the symmetry broken state is hard to know from theory and will depend on the specific symmetries imposed. Furthermore, to understand our experimental data ($dn_D/dn$) we are mainly interested in the evolution of $\varepsilon_F$ which is not affected significantly by the exact form of the symmetry broken state at $\nu = 2$, but rather by the energy shift. To this end, we keep our analysis general and use as an ansatz a simple Stoner instability model where for $2 \leq \nu \leq 2.6$ the FB of one flavor type is increased in energy by $\Delta_{HF} = 20$ meV and the other is reduced by $\Delta_{HF}$ (Figs. 3f,g). Flavor here can include the Chern basis, intervalley coherent, or involve the graphene sublattice. No matter the actual state, the broad features will be similar and the important energy shift $\Delta_{HF}$ will appear.

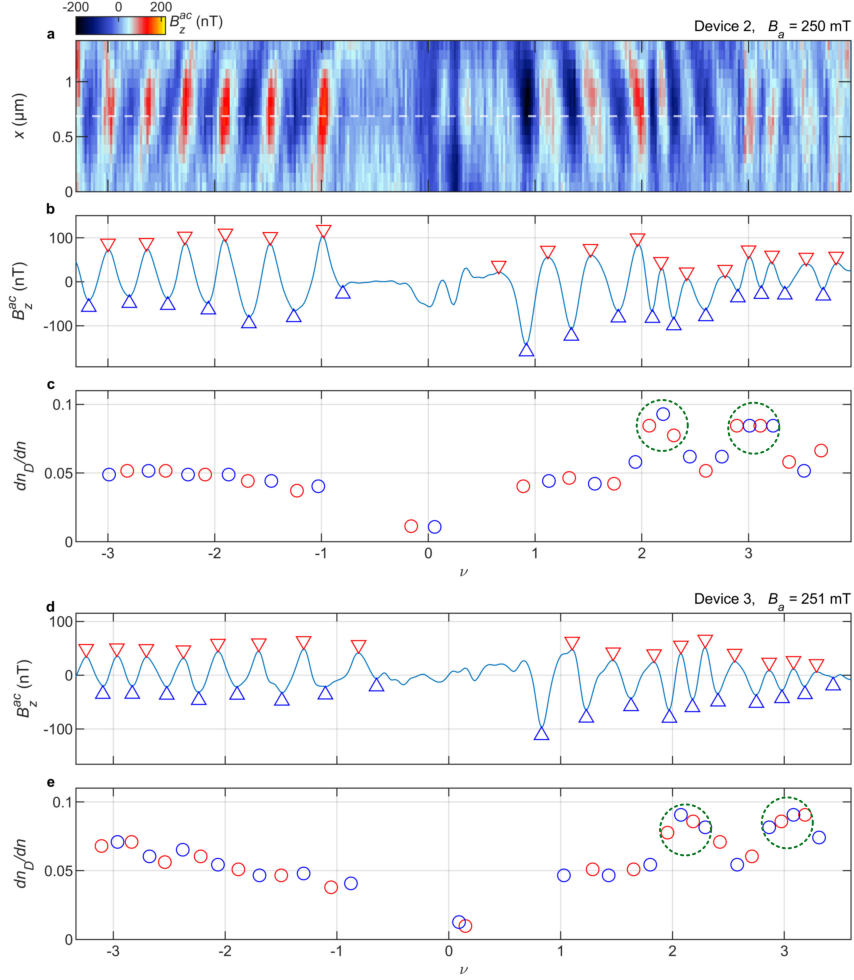

**Extended Data Fig. 2. QOs in Devices 2 and 3 ($\theta \approx 1.5°$). a,** $B_z^{ac}(x, \nu)$ in Device 2 at $D = 0$ and $B_a = 250$ mT upon hole and electron doping. **b,** Cross-section of $B_z^{ac}(\nu)$ along the white dashed line in (a). **c,** Fraction of carriers added to the system that go to the Dirac sector, $dn_D/dn$ vs. doping, extracted from (b). Green circles highlight peaks in $dn_D/dn$ near $\nu = 2$ and $3$ indicating symmetry breaking. **d,** $B_z^{ac}(\nu)$ in Device 3. **e,** $dn_D/dn$ vs. doping extracted from (d). Green circles highlight peaks in $dn_D/dn$ near $\nu = 2$ and $3$.

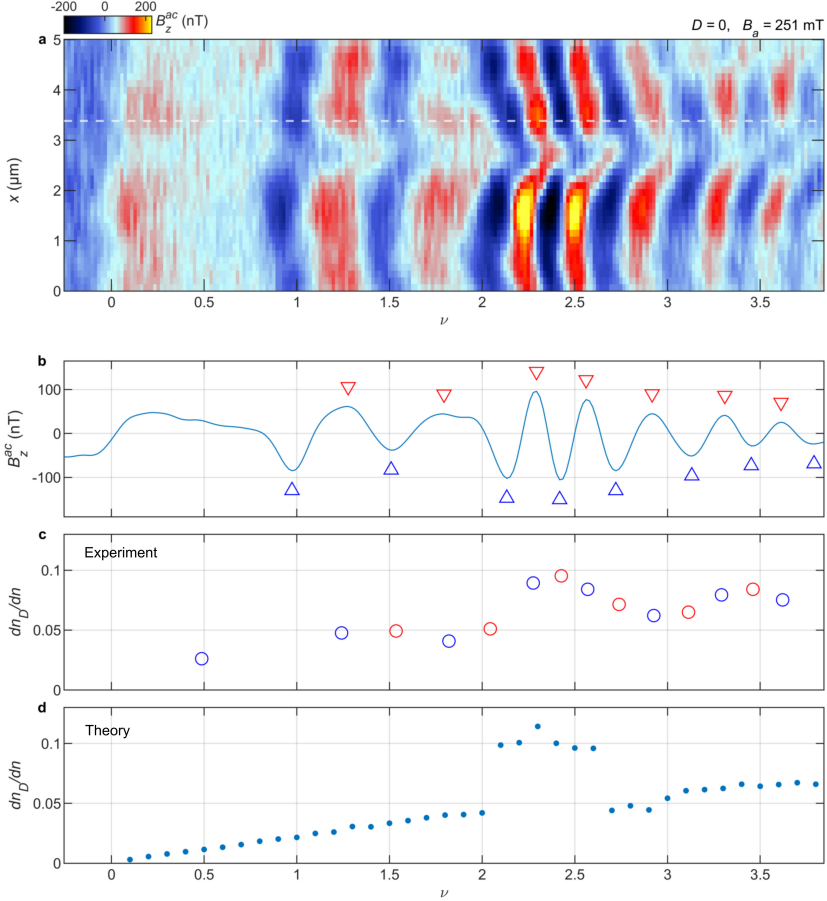

**Extended Data Fig. 3. QOs in Devices 1 upon electron doping at $B_a = 251$ mT. a-d,** Same as in Figs. 3a-d but at a higher $B_a = 251$ mT. **a,** $B_z^{ac}(x, \nu)$ upon electron doping. **b,** Cross-section of $B_z^{ac}(\nu)$ along the white dashed line in (a). **c,** $dn_D/dn$, extracted from (b) showing a peak above $\nu = 2$, indicating degeneracy lifting. The resolution of the peak is poorer than in Fig. 3c due to sparser LLs at higher $B_a$. **d,** $dn_D/dn$ from the Hartree calculation.

## Theoretical modeling of band renormalization in tTLG

### I. Continuum model

We begin by recapitulating the calculation of the non-interacting band structure of tTLG, expanding on the treatment of the Bistritzer-MacDonald (BM) model [19,65]. As usual, we restrict ourselves to calculation of the band structure near the $K$ valley of monolayer graphene, since the bands near $K'$ can be recovered by applying time-reversal symmetry to the $K$ bands. The Hamiltonian of the three decoupled layers expanded around the $K$ valley is

$$H_d = \sum_{l=1,2,3} \sum_{\boldsymbol{k}} \psi_{l,\boldsymbol{k}}^{\dagger} h_{l,\boldsymbol{k}} \psi_{l,\boldsymbol{k}}, \tag{2}$$

where $\psi_{l,\boldsymbol{k}}$ is a spinor in sublattice space of electronic annihilation operators on layer $l$ at momentum $\boldsymbol{k}$ (relative to $K$), the intralayer Hamiltonian is $h_{l,\boldsymbol{k}} = v_F e^{-\frac{1}{2}i\theta_l \sigma_z} \left( k_x \sigma_x + k_y \sigma_y \right) e^{\frac{1}{2}i\theta_l \sigma_z} + (\frac{l}{2} - 1)U$, $\theta_l$ is the twist angle applied to the $l$ layer, $U$ is the potential difference between the outermost layers, $v_F$ is the monolayer graphene Fermi velocity, and the $\sigma_i$ Pauli operators act on the sublattice degree of freedom. In the alternating-angle tTLG case, $\theta_1 = \theta_3 = \frac{\theta}{2}, \theta_2 = -\frac{\theta}{2}$, and we refer to $\theta$ as the relevant twist angle.

The interlayer tunneling is described by (neglecting tunneling between the outermost layers)

$$H_t = \sum_j \sum_{\boldsymbol{k}} (\psi_{1,\boldsymbol{k}}^\dagger + \psi_{3,\boldsymbol{k}}^\dagger) T_j \psi_{2,\boldsymbol{k}+q_j} + h.c., \tag{3}$$

where the tunneling matrices are given by

$$T_j = \begin{pmatrix} w_0 & w_1 e^{-\frac{2\pi}{3}i\,(j-1)} \\ w_1 e^{\frac{2\pi}{3}i\,(j-1)} & w_0 \end{pmatrix} \tag{4}$$

and $\boldsymbol{q}_1 = L\hat{y}, \boldsymbol{q}_2 = \frac{L}{2}(-\sqrt{3}\hat{x} - \hat{y}), \boldsymbol{q}_3 = \frac{L}{2}(\sqrt{3}\hat{x} - \hat{y}), L = \frac{8\pi}{3a}\sin\frac{\theta}{2}$. Here $w_0$ and $w_1$ are the tunneling strengths in the AAA stacked and ABA stacked regions, respectively. Due to lattice relaxation, the AAA regions shrink, and thus $w_0$ is phenomenologically lowered in comparison to $w_1$.

First, we focus on a system absent of an electric displacement field. In this scenario a mirror symmetry exists and the intralayer Hamiltonians for layers $l = 1$ and 3 are identical. It is then instructive to re-define creation and annihilation operators in a new basis, $\psi_{\pm,\boldsymbol{k}} \equiv \frac{1}{\sqrt{2}}(\psi_{1,\boldsymbol{k}} \pm \psi_{3,\boldsymbol{k}})$. The $H_d$ part of the Hamiltonian remains decoupled in this new basis, and the interlayer part $H_t$ only couples $\psi_{+,\boldsymbol{k}}$ to $\psi_{2,\boldsymbol{k}}$, with an extra factor of $\sqrt{2}$ which enters the tunneling matrix. The Hamiltonian thus is decomposed into two decoupled sectors: a Dirac-cone Hamiltonian with velocity $v_F$, and a Bistritzer-Macdonald FB sector with a modified magic angle $\theta_{tTLG} = \sqrt{2}\theta_{TBG}$ (where $\theta_{TBG}$ is the magic angle of twisted bilayer graphene [65]).

Taking into account the moiré periodicity of the continuum model, it is useful to shift the momentum origin of the middle $l = 2$ layer by $\boldsymbol{q}_1$, and define the moiré lattice vectors $\boldsymbol{b}_1 = \boldsymbol{q}_3 - \boldsymbol{q}_1$, and $\boldsymbol{b}_2 = \boldsymbol{q}_3 - \boldsymbol{q}_2$. We may then write the total non-interacting Hamiltonian $H_0 = H_d + H_t$ as

$$H_0 = \sum_{l=1,2,3} \sum_{\boldsymbol{G}} \sum_{\boldsymbol{k}} \psi_{l,\boldsymbol{k}+\boldsymbol{G}}^\dagger h_{l,\boldsymbol{k}+\boldsymbol{G}} \psi_{l,\boldsymbol{k}+\boldsymbol{G}} + \sum_j \sum_{\boldsymbol{G},\boldsymbol{G}'} \sum_{\boldsymbol{k}} f_{j;\boldsymbol{G},\boldsymbol{G}'} (\psi_{1,\boldsymbol{k}+\boldsymbol{G}}^\dagger + \psi_{3,\boldsymbol{k}+\boldsymbol{G}}^\dagger) T_j \psi_{2,\boldsymbol{k}+q_j+\boldsymbol{G}'}, \tag{5}$$

where $\boldsymbol{G}, \boldsymbol{G}'$ are reciprocal lattice vector spanned by $\boldsymbol{b}_1, \boldsymbol{b}_2$, and $f_{j;\boldsymbol{G},\boldsymbol{G}'} = 0,1$ defines the nearest-neighbor-only connectivity between Dirac cones in momentum space. Notice that in Eq. (5) the momentum $\boldsymbol{k}$ summation is only over the mBz. In order to recover the band structure, we diagonalize $H_0$. We find that for the BS of lowest-lying bands to converge, it is sufficient to keep only reciprocal lattice vectors at size $6|\boldsymbol{G}|$.

## II.  Interactions and Hartree potential

To recover the interaction induced BS modifications we now need to take the Coulomb repulsion between electrons into account. This is done in the usual way, by considering:

$$H_I = \frac{1}{2} \int d\boldsymbol{r} d\boldsymbol{r}'\, \bar{\rho}(\boldsymbol{r}) V(\boldsymbol{r} - \boldsymbol{r}') \bar{\rho}(\boldsymbol{r}'), \tag{6}$$

where $\bar{\rho}(\boldsymbol{r})$ is the density operator relative to charge neutrality, and the Coulomb potential is $V(\boldsymbol{r}) = e^2/4\pi\epsilon|\boldsymbol{r}|$ ($\epsilon$ is the effective dielectric constant). The contribution that is most relevant to our experimental results, i.e., the relative density dependent shift in energy between the Dirac band and the FBs, can be well described by resorting to a Hartree approximation of the Coulomb interaction term. As a result, the band structure will be determined by self-consistent calculation of the mean field Hamiltonian,

$$H_{MF} = H_0 + H_{Hartree}(\nu),$$

(7)

where the filling-dependent Hartree term is given by

$$H_{Hartree} = \frac{1}{\Omega} \sum_{k,k',q} V_q \langle \psi^{\dagger}_{l',k'-q} \psi_{l',k'} \rangle_{\nu} \; \psi^{\dagger}_{l,k+q} \psi_{l,k}.$$

(8)

Here a summation over repeated layer indices is assumed, $V_q$ is the Fourier transform of $V(\boldsymbol{r})$, and $\Omega$ is the system volume. The expectation value $\langle \, . \, \rangle_{\nu}$ is evaluated by finding the ground state of $H_{MF}$, $|\Psi_{\nu}\rangle$, and calculating $\langle \, . \, \rangle_{\nu} = \langle \Psi_{\nu}| \, . \, |\Psi_{\nu}\rangle - \langle \Psi_0 | \, . \, |\Psi_0\rangle$, i.e., relative to charge neutrality.

The $\boldsymbol{q} = 0$ part of this expectation value is exactly offset by the contribution of the background charge, and is thus discarded. Assuming momentum conservation, that would have been the end of the story. However, due to the moiré pattern which breaks the translation symmetry on the graphene scale, momentum is only conserved up to some reciprocal lattice vector $\boldsymbol{G}$. In principle, one should consider all possible $\boldsymbol{q} = \boldsymbol{G}$ contributions, yet it has been shown [55] that only the first "star" of reciprocal lattice vectors $|\boldsymbol{G}| = \sqrt{3}L$ significantly contribute to the Hartree term. Thus, we write

$$H_{Hartree} = \frac{1}{\Omega} {\sum_{\boldsymbol{G}}}' V_G \sum_{k'} \langle \psi^{\dagger}_{l',k'-\boldsymbol{G}} \psi_{l',k'} \rangle_{\nu} \sum_{k} \psi^{\dagger}_{l,k+\boldsymbol{G}} \psi_{l,k},$$

(9)

where the summation ${\sum_{\boldsymbol{G}}}'$ is over the six reciprocal lattice vectors in the first star. The value of the Coulomb repulsion at these momenta is $V_G = e^2 / (\epsilon a / 2 \sin(\frac{\theta}{2}))$. Further noticing the fact that $V_G \sum_{k'} \langle \psi^{\dagger}_{l',k'-\boldsymbol{G}} \psi_{l',k'} \rangle_{\nu}$ should all be equal within this first star, we define

$$V_H \equiv \frac{1}{6\Omega} {\sum_{G}}' V_G \sum_{k'} \langle \psi^{\dagger}_{l',k'-\boldsymbol{G}} \psi_{l',k'} \rangle_{\nu},$$

(10)

and finally write Eq. (9) in the concise form:

$$H_{Hartree} = V_H {\sum_{\boldsymbol{G}}}' \sum_{k'} \psi^{\dagger}_{l,k+\boldsymbol{G}} \psi_{l,k}.$$

(11)

We note that the form of Eq. (11) suggests a quite simple interpretation of the Hartree band renormalization term: it is a periodic potential (with the moiré pattern periodicity) acting on the electrons by the spatially non-uniform distribution of the same electrons.

To find the value of the $\nu$-dependent Hartree potential we employ an iterative self-consistent method. An initial value $V_H^{(0)}$ is chosen. The mean-field Hamiltonian $H_{MF}$ [Eq. (7)] is then diagonalized using this initial value. We then use the ground state of the mean-field Hamiltonian to calculate the next value $V_H^{(1)}$ as in the expression above Eq. (10). This step would depend explicitly on the specific filling factor $\nu$ at which we calculate $V_H$. We repeat these steps until the value of $V_H$ converges, which typically takes no more than $\sim$ 10 steps.

### III.  Effect of displacement field

The presence of a displacement field induces an interlayer potential, $U \neq 0$, between $l = 1$ and $3$. This couples the $\psi_{\pm}$ sectors through a term $\sum_{\boldsymbol{k}} \psi_{+,\boldsymbol{k}} \left(-\frac{U}{2}\right) \psi_{-,\boldsymbol{k}} + h.c.$, which in turn couples the FB sector with the Dirac cone sector. As a result, the FBs and the Dirac cone hybridize at the Dirac point. $V_{Hartree}$ is calculated similarly in a self-consistent fashion with $U \neq 0$. Extended Data Figs. 4a-e show the renormalized Hartree potential BS for several values of $U$. As $U$ is increased the graphene and FB Dirac cones hybridize and are pushed to higher energy. Subsequently, $\nu$ must be increased to larger values before $\varepsilon_F$ reaches the Dirac node (red arrows), where the approximate $N_D = 0$ LL appears. This causes the general trend of Dirac LLs dispersing to higher $\nu$ as a function of $U$. The same effect occurs in the single particle picture (Extended Data Fig. 4f), and more generally in any model where the graphene and FB Dirac nodes overlap in $k$-space, namely $C_3$ is not broken or FB Dirac cones are not gapped, as discussed in the following section.

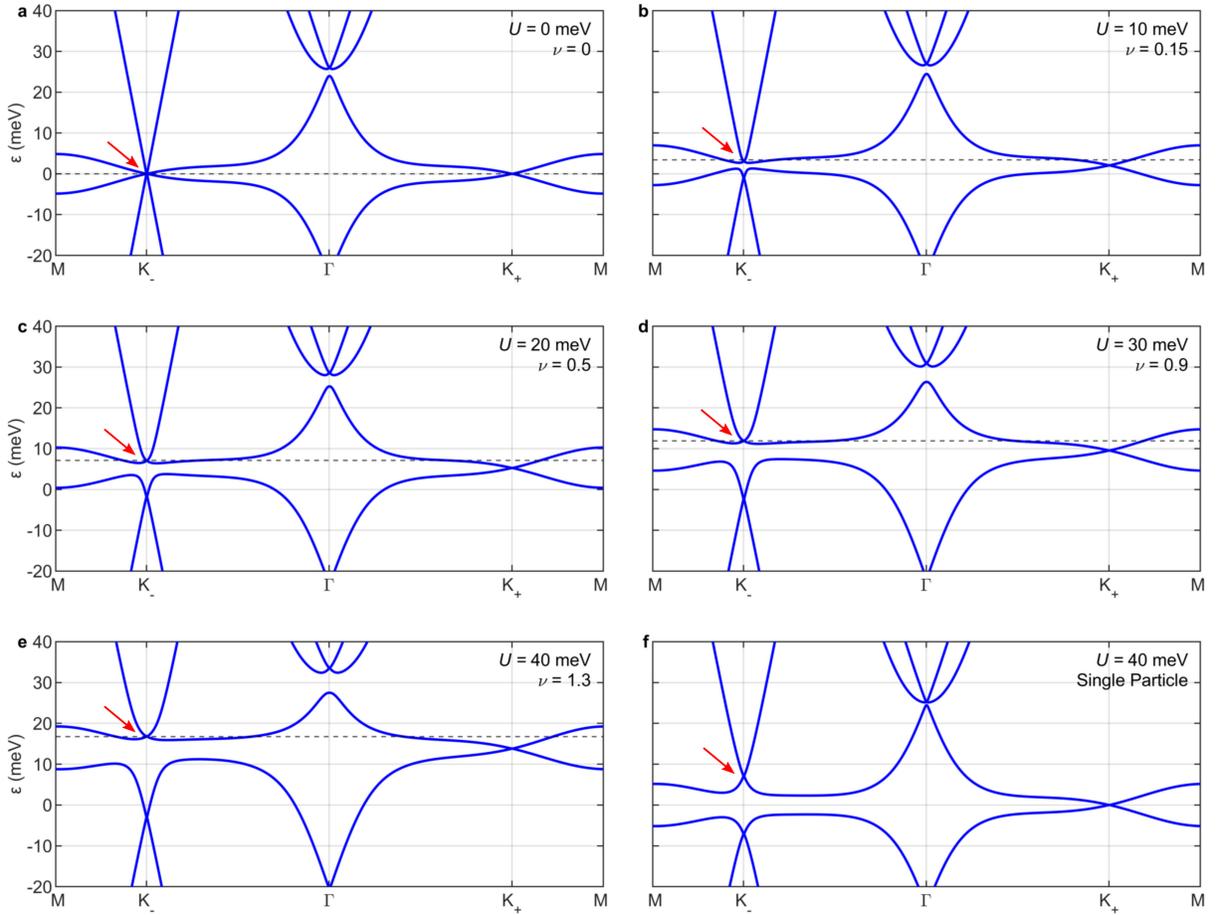

**Extended Data Fig. 4. Hartree potential band structure with $U \neq 0$. a-e,** Band structure line cuts for a single valley with self-consistently calculated Hartree potential for $U = 0$ (**a**), $U = 10$ (**b**), $U = 20$ (**c**), $U = 30$ (**d**), and $U = 40$ meV (**e**). Band structure cuts are shown at the indicated values of $\nu$ for which $\varepsilon_F$ (dashed line) reaches the hybridized Dirac point where the $0^{\text{th}}$ LL resides, illustrating that at larger $U$ Dirac LLs appear at larger $\nu$ as shown in Fig. 4c. **f,** Band structure line cut for $U = 40$ meV for a single particle calculation. Here, a similar effect is seen where the graphene and FB Dirac cones overlap and hybridize at zero energy, pushing all $N^D$ LLs up in energy.

To calculate the Dirac LL spectrum at $B_a = 251$ mT, as seen in Fig. 4c (grey dashed), we track the density of carriers in the Dirac cone, $n_D$, as a function of $\nu$. At values of $\nu$ for which $n_D = 4N_D B_a / \phi_0$, the compressible $N_D$ LL is half filled corresponding to the peaks in QOs. This procedure is done for a number of values of $U$, and the results interpolated and smoothed to get the continuous spectrum shown in Fig. 4c (grey dashed).

As discussed in the main text, this Hartree only calculation is sufficient to describe the experimental observations for $|\nu| \gtrsim 1.5$, but to explain the low energy physics near CNP, inclusion of the Fock term is necessary.

### I.   Full Hartree Fock calculations

In this section we describe our self-consistent Hartree Fock calculations of the low energy tTLG bands [23,27,30,33]. We use a dual-gate screened Coulomb interaction with screening length $d_s = 20$ nm and effective dielectric constant $\epsilon \approx 16 - 17$ that accounts for both hBN screening and screening from remote tTLG electrons [20], the latter modeled through static Dirac cone screening. We further use the infinite temperature subtraction scheme to address double counting of interaction effects in band-structure parameters [26,66]. The specific BS parameters used are described in the following section, although we have checked that our qualitative conclusions are robust to varying these parameters within a physically reasonable range.

The inclusion of the Fock term is crucial for symmetry breaking states. While for symmetric states the Fock dispersion only acts to renormalize the non-interacting band structure, for symmetry breaking states the dispersion is drastically altered. Previously, for $\nu = 2$, we approximated Hartree Fock dispersion of Stoner states by splitting the Hartree bands symmetrically with a phenomenological $k$-independent Fock energy $\Delta_{HF}$. This is a reasonable approximation at the Fermi momenta, which are not too close to the $\Gamma$ point at $\nu = 2$, such that the $k$-dependence of the Fock dispersion is weak. In contrast, near charge neutrality, we will see that the large scale and strong $k$-dependence of the Fock Hamiltonian makes it the dominant term; the Hartree term vanishes at charge neutrality, and the Fock term dwarfs the non-interacting band structure. Furthermore, the low energy dispersion of the Fock term is near $\Gamma$ for all symmetry breaking states.

While one could, with significant effort, perform self-consistent Hartree Fock across all filling factors, it is a much more controlled approximation at integer filling due to the proximity to the strong-coupling limit, and it does not include, for example, the dynamical screening of flat-band electrons that is likely necessary for a quantitatively correct treatment of the metallic states. We will therefore focus our efforts at charge neutrality, where the Fock term is most needed and most reliable.

Our goal is to investigate the influence of various charge-neutrality symmetry breaking states on the graphene Dirac cone and its LLs through self-consistent Hartree Fock. The results we find are consistent with prior works [13,20,25]. For simplicity we assume spinless electrons and we take $N = 4$ bands per valley which is the minimal number of bands needed to capture the FBs and the low-energy part of the decoupled graphene Dirac cone. While all simulations were first done keeping both valleys, for the valley-diagonal NSM state we subsequently focused on a single valley in order to increase $k$-space grid density.

By imposing or not imposing each of $U(1)_V$ (valley conservation) and $C_2 T$ (two-dimensional inversion combined with time-reversal) symmetries, we can disallow or allow for different types of symmetry breaking states. If no symmetries are imposed, we obtain the KIVC state [23] at not too large displacement field $D$, as expected from strong coupling arguments [20,23,24,29,31]. The KIVC state, breaks valley conservation causing a folding of both valleys into one mBz, preserves $C_2 T$, and gaps the FB Dirac cones through intervalley hybridization (Extended Data Fig. 5c). If we impose $U(1)_V$, ruling out KIVC, we either obtain a $C_3$ breaking NSM [30], or a VH state which breaks $C_2T$. The VH state gaps the graphene Dirac cone with a mass $\propto U^2$. This is inconsistent with experiment, as the graphene Dirac cone $0^{\text{th}}$ LL does not split and disperse away from CNP with $D$ field (Fig. 4a,b). We will therefore not consider the VH state further.

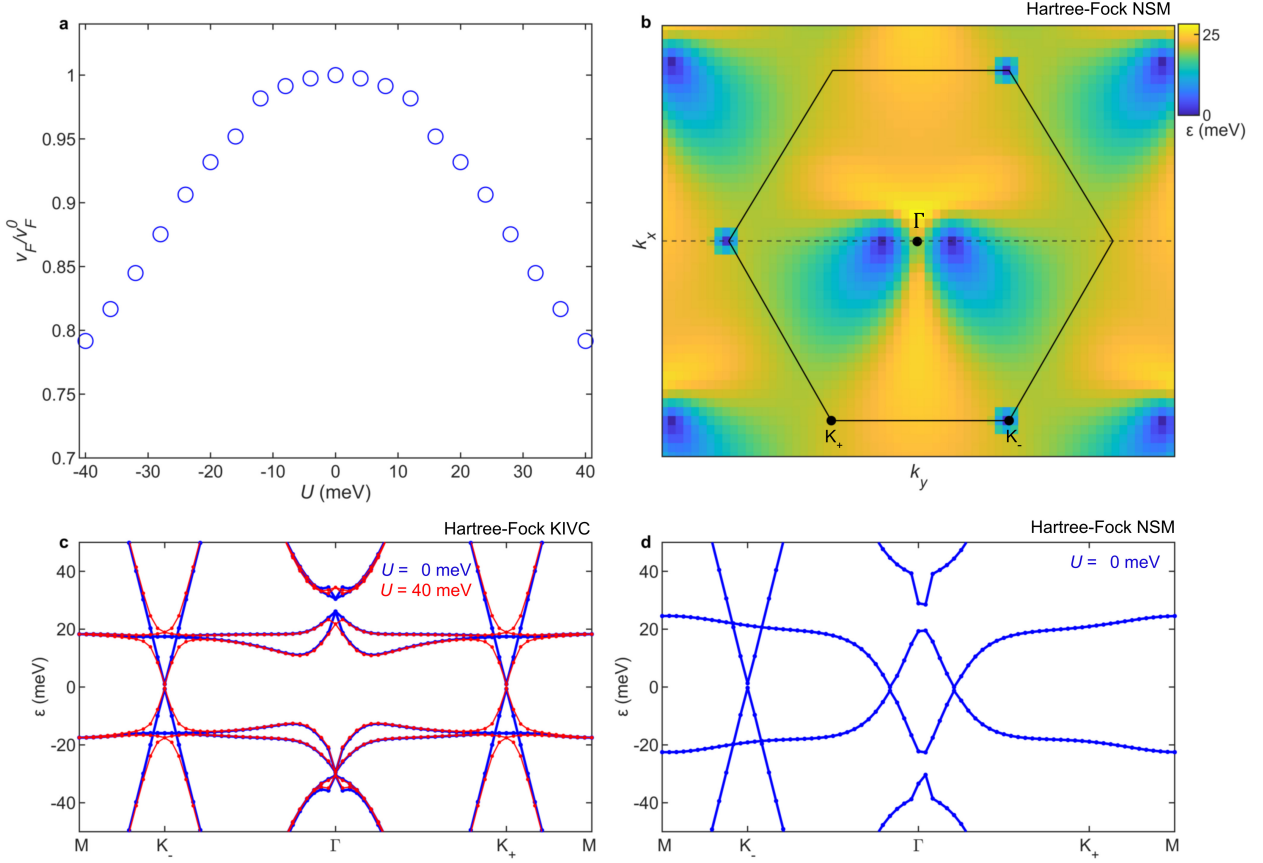

**Extended Data Fig. 5. Hartree Fock calculations. a,** Calculation of the renormalization of the Dirac cone $v_F$ at CNP as a function of $U$ in the NSM state. Values are normalized by $v_F^0$, $v_F$ at $U = 0$, and used to plot the LL spectrum in Fig. 4c. **b,** BS of the conduction FB, $\varepsilon(k_x, k_y)$, of the HF NSM state at $U = 0$ in an extended mBz for the $K$ valley ($K'$ is related by a 180° rotation). The system spontaneously breaks $C_3$ symmetry and FB Dirac cones migrate towards $\Gamma$, while the graphene Dirac point stay at the mBz corner. The dashed line marks the linecut along which (d) and Fig. 4d are plotted. **c,** BS of the gapped KIVC state for $U = 0$ (blue) and $U = 40$ meV (red). Coherence between $K$ and $K'$ folds both bands into a single mBz. Application of $U$ hybridizes the bands and renormalizes the Dirac cone $v_F$, similar to the NSM state. **d,** BS of the NSM state at $U = 0$ along the back dashed line in (b). The system spontaneously breaks $C_3$ and the FB Dirac cones are shifted towards $\Gamma$.

Nominally the NSM is disfavored relative to KIVC as well, but it becomes the ground state in the presence of a small amount of strain [30,51]. Because $C_2T$ and $U(1)_V$ is preserved, and the two FB Dirac cones have the same chirality, the Dirac cones cannot gap out. In a three-fold symmetric state, the FB Dirac cones are further pinned to the moiré $K$-points leading to immediate $D$-field induced hybridization with the graphene Dirac cone, as shown in the previous section when only considering the Hartree potential. However, the $C_3$ breaking of the NSM state moves the FB Dirac cones to a close vicinity of the $\Gamma$ point (Extended Data Fig.

5b). At the $K$-point, the graphene Dirac cone is protected by a FB $K$-gap and $C_2T$ symmetry, similar to KIVC state. After adding a displacement field, the graphene Dirac cone shifts slightly, and its velocity becomes slightly anisotropic, as expected since it begins to perturbatively feel the $C_3$ breaking of the moiré electrons. However, it remains at zero energy [13,20,25] and the hybridization occurs at $\Delta_{HF}$ in sharp contrast to the immediate hybridization in the $C_3$ symmetric case. With increasing $U$, the hybridization increases and the graphene Dirac cone $v_F$ is effectively renormalized (Fig. 4d). A similar effect happens in the KIVC state (Extended Data Fig. 5c).

To quantify this effect and facilitate qualitative comparison to experiment, we evaluate the NSM BS at various values of $U$. Memory constraints in HF simulations allow only sparse $k$-space grids ($36x41$), especially when considering a dispersive Dirac cone. To this end we interpolate the subsequent 3D BS and take FS cuts of the renormalized Dirac cone. $v_F$ is then computed with respect to the symmetric case at $U = 0$ (Extended Data Fig. 5a), and then the $N_D = \pm1$ LL is scaled as a function of $U$ by the renormalized $v_F$ (Fig. 4c, magenta).

Note that in the NSM the migration of the FB Dirac cones to $\Gamma$ is due to the concentration of the FB Berry curvature (in the Chern-basis [23]). Roughly speaking, the Dirac cones behave like $k$-space vortices that see the Chern-basis Berry curvature as a $k$-space magnetic field. The energetic competition between the repulsion of vortices and the tendency of a vortex to maximally overlap the magnetic fields, then determines the final location of the FB Dirac cones [20,30,67]. The uniformity of the Berry curvature in the FBs is mostly governed by the gap at the $\Gamma$ point to the dispersive bands. At zero gap, the Berry curvature is a delta function and the FB Dirac cones squeeze close to the $\Gamma$ point with larger FB $v_F$, whereas if the gap is increased the Berry curvature is more spread out and the FB Dirac cones reside further from the $\Gamma$ point with lower FB $v_F$. Therefore the $\Gamma$ point gap size determines the low density FB Dirac cone $v_F$, which is what governs $dn_D/dn$, the quantity compared to experiment (Fig. 4f,g).

## II.    Lack of gap at CNP

Experiment shows a significant density in the FB sector below the first peak in magnetization due to equilibrium currents flowing in the incompressible gap between the $N_D = 0$ and 1 LLs. At 56 mT (Fig. 4b), this can be approximated to be at an energy equal to $\varepsilon_1^D/2 = 3.9$ meV, implying a significant FB density below such a small energy.

As discussed in the main text, gaped states should generically be of order of the Coulomb scale, $\sim20$ meV. This, coupled with the fact that the $D$ field dependance of the local LLs (Figs. 4a,b) give a $K$-point gap of the theoretically expected order ($\pm19$ meV), is strong evidence of the lack of a gap at CNP in our device.

Additionally, other local probes, such as STM have measured the gap at $\nu = 2$ in MATBG of order 10 meV [36], which is likely an underestimate due to the metallic screening of the STM tip. This is again in agreement with the theoretical order. Although some transport measurements at $\nu = 2$ measure smaller gaps (~few meV), this should be considered as a lower bound due to averaging over μm scale in systems that have large disorder, which a local probe avoids.

To put a more stringent bound on a possible gap in the system, we preform simulations of the QOs in magnetization on a simplified model and compare to experimental $B_z^{qc}(n)$ (Extended Data Fig. 6). To model the FB at low densities we take two Dirac cones with renormalized $v_F^{FB} \approx v_F/3.5$, and calculate the QOs arising from the LLs in the Dirac sector broadened by the Dingle $\Gamma \cong 1$ meV, as described in [44]. We note that to best match the experiment we need to add some small mass to the FB cones, such that the zero crossing is not linear. This implies an enhanced DOS in the FB at CNP that can be explained either by disorder or by a NSM state in which the FB Dirac cones move very close to each other such that they hybridize and effectively have a parabolic dispersion at very low energy. Finally, we allow a gap $\Delta$ to open in the FB and extract $\partial M / \partial n \propto B_z^{ac}$ by convolving the calculated magnetization with our *ac* density excitation, $n^{ac}$. With increasing $\Delta$, a sharp peak appears at $n \approx 0.04 \times 10^{12}$ (Extended Data Fig. 6b), which is not present in experiment. Even at $\Delta = 1$ meV (red), a small peak is discerned in simulations. Therefore, we place as a conservative estimate, $\Delta < 1$ meV. Note that the steep initial rise of the calculated $\partial M / \partial n$ cresting at the peak, in contrast to the much more gradual initial increase in the experimental $B_z^{ac}$, is the result of rapid initial filling of the Dirac band for finite $\Delta$ due to the absence of DOS in the FB.

A very small gap of 1 meV or smaller cannot be completely ruled out experimentally. Indeed, there can be a second order phase transition between the NSM and KIVC state, tuned by the strength of the strain [51]. In this scenario, the state would have both a NSM order parameter (which is $C_3$ breaking) and a KIVC order parameter, where the NSM order parameter is much larger than the KIVC. Effectively this can be thought of as an NSM that breaks $C_3$ with a small mass term induced by the small KIVC order parameter. Therefore, even if there were to be a small FB gap that experiment is not sensitive to, the state should still be thought of as a NSM from a theoretical perspective.

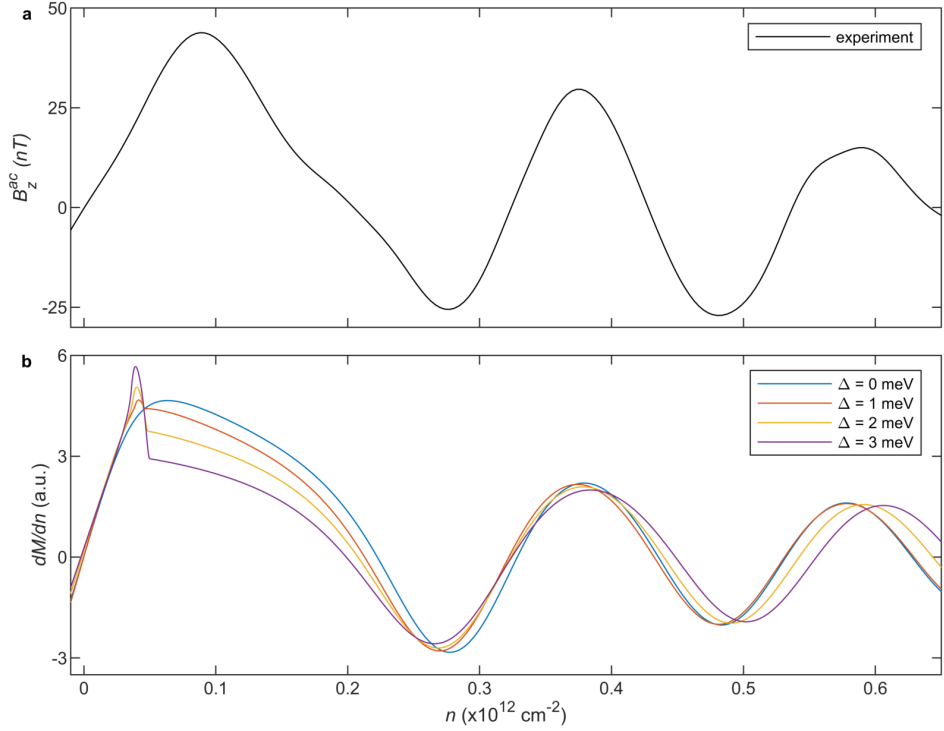

**Extended Data Fig. 6. QO simulations at CNP. a,** Zoomed-in $B_z^{ac}(n)$ near CNP at $B_a = 56$ mT, reproduced from Fig. 4e. **b,** Simulations of QOs, $\partial M / \partial n$, arising from the Dirac band in the presence of FB described by Dirac dispersion with reduced $v_F^{FB}$ and variable FB gap $\Delta$ from 0 to 3 meV. Even a small gap of $\Delta = 1$ meV is inconsistent with the shape of the measured QOs in (a).

### III. Fitting BS parameters

The $B_a = 251$ mT experimental data and derived $dn_D / dn$ in Fig. 2 allow for sensitive fitting of BS parameters [43,44] with the information from the FBs and up through the dispersive bands ($\nu > 4.2$). Particularly, there are three features that we would like to fit:

a) The maximal value of $dn_D / dn \approx 0.1$ at the top of the FB.

b) The lack of a large gap to the dispersive band in the FB sector. A large gap would result in $dn_D/dn = 1$ in the gap, since there would be only the graphene Dirac cone Fermi surface, which we do not observe.

c) Increased $dn_D/dn \approx 0.2$ in the dispersive bands.

Additionally, we have the $B_a = 56$ mT experimental data and derived $dn_D/dn$ in Fig. 4f with information about the physics near CNP. As described in the previous section, the gap to the dispersive band at $\nu = 0$ is related to $v_F$ of the FB Dirac cones in the NSM state, and therefore to the DOS of the FBs near CNP. We find that with no gap, $dn_D/dn$ is too large (~0.1). Therefore, to fit both high and low field data (or equivalently, both high and low energy physics), we need a clear single particle gap to the dispersive bands at $\nu = 0$, that is not too large when reaching $\nu \approx 4$ from the added Hartree potential (Figs. 2h-j).

We assign the generally accepted value $\gamma_0 = 2800$ as a starting point and span the generally expected range of the other parameters. We find that using $w_1$ = 110 meV, $w_0 = 0.72w_1$, and $\epsilon \approx 17$ matches both the high and low energy physics. Due to the large number of parameters compared to the number of features, there can be other parameter sets that would match qualitatively our data.

Note that it would not be surprising if parameters that yield zero gap at $\nu \approx 4$ give only a qualitatively accurate DOS at charge neutrality. There are many effects in practice that the BM model does not take into account, such as $C_3$ symmetric strain due to lattice relaxation, as well as heterostrain. Additionally, there is no source of particle-hole symmetry breaking. These could have distinct quantitative effects, that are hard to predict on these two very different states – NSM at charge neutrality and the Hartree renormalized BM bands at $\nu \approx 4$. Furthermore, the sparse $k$-space grid for HF simulations leads to inherent numerical uncertainty and should only be taken qualitatively. For these reasons, qualitative agreement is the realistic measure in FB graphene systems. We are not aware of any works that have good quantitative agreement across such different filling factors for the same parameters.

**Supplementary Movie 1 | Doping dependent band structure**. Self-consistent calculation of the mean field Hartree interaction term which leads to a doping dependent band structure. At $\nu = 0$ the Hartree interaction term is zero and the BS is equivalent to a single particle continuum model calculation. As $\nu$ is increased the Hartree term increases and states away from $\Gamma$ have their energy increased. The dashed gray line marks $\varepsilon_F$, which increases at a much faster rate. Red dashed lines correspond to the Dirac Landau levels at $B_a = 251$ mT, as in Fig. 2.